\documentclass[fleqn,usenatbib]{mnras}
\UseRawInputEncoding
\usepackage{xcolor}
\usepackage{newtxtext,newtxmath}
 
\usepackage[T1]{fontenc}
\usepackage{lscape}
\usepackage[normalem]{ulem} 
\usepackage{natbib}

\DeclareRobustCommand{\VAN}[3]{#2}
\let\VANthebibliography\thebibliography
\def\thebibliography{\DeclareRobustCommand{\VAN}[3]{##3}\VANthebibliography}

\usepackage{graphicx}	% Including figure files
\usepackage{amssymb}% Extra maths symbols
\usepackage{amsmath}	% Advanced maths commands
\usepackage{threeparttable}

\definecolor{address}{rgb}{0.36,0.54,0.66}
\definecolor{red}{rgb}{0.8,0.,0.}

\newcommand{\HI}{\mathrm{H}\,\textsc{\large{i}}}
\newcommand{\HIi}{H\,\textsc{\large{i}}}
\newcommand{\HIs}{\mathrm{H}\,\textsc{\textmd{i}}}
\newcommand{\HIb}{\mathbf{H}\,\textsc{\large{i}}}
\newcommand{\HIlb}{\mathbf{H}\,\textsc{\Large{i}}}

\newcommand{\Hmol}{\mathrm{H_2}}
\newcommand{\HIl}{\mathrm{H}\,\textsc{\Large{i}}}

\newcommand{\HIhi}{H\,\textsc{\huge{i}}}
\newcommand{\kms}{\mathrm{km\,s}^{-1}}
\newcommand{\msol}{\mathrm{M}_\odot}
\newcommand{\mstar}{M_\star}
\newcommand{\K}{\,\mathrm{K}}
\newcommand{\HIn}{H\,{\sc{\Large i}}{ }}
\usepackage{orcidlink}

\graphicspath{{Figures/}}

\title[$\HIhi$ line asymmetries of isolated galaxies]{Asymmetries in spatially unresolved 21-cm emission line profiles of isolated galaxies} 

\author[A. Manuwal et al.]{
\parbox{17cm}{
Aditya Manuwal$^{\orcidlink{0000-0003-2893-2793}}$,$^1$\thanks{E-mail: adi.manuwal@astro.unam.mx}
H. M. Hern\'andez-Toledo,$^1$
J. A. V\'azquez-Mata$^{\orcidlink{0000-0001-8694-1204}}$,$^1$
Diana Serrato Barrajas$^1$,
Aldo Rodr\'iguez-Puebla$^{\orcidlink{0000-0002-0170-5358}}$,$^{1}$
Vladimir \'Avila-Reese$^{\orcidlink{0000-0002-3461-2342}\,1}$ and
A. R. Calette$^{\orcidlink{0000-0001-8798-2542}\,1}$
}
\vspace{0.3cm}
\\
$^{1}$Universidad Nacional Aut\'onoma de M\'exico, Instituto de Astronom\'{\i}a, A.P. 70-264, 04510 CDMX, M\'exico\\
}

\date{Accepted 2026 July 17. Received 2026 June 30; in original form 2026 March 2}

\pubyear{2026}

\begin{document}
\label{firstpage}
\pagerange{\pageref{firstpage}--\pageref{lastpage}}
\maketitle

\begin{abstract}
The origin of asymmetry in the $\HIl$ of galaxies remains an elusive problem,
largely due to the difficulties associated with distinguishing between its secular and external contributors. We have compiled a sample of local isolated galaxies from the UNAM-KIAS and the latest AMIGA samples with near-complete beam coverage and robust estimates of $\HIl$ line asymmetry. The $\HIl$ measurements are based on single-dish spectra sourced from five surveys: $\HIl$-MaNGA, NIBLES, KLUN, ALFALFA, and xGASS. Our galaxies tend to be late-type, star-forming centrals at all masses but exhibit slightly lower specific star formation rates (sSFRs) and higher $\HIl$ contents than expected. The latter is likely driven by lower star formation efficiency, weaker outflows, and additionally, higher accretion rate of gas onto the galaxy at $\mstar\lesssim 10^{10.3}\,\msol$. AMIGA, however, shows systematically higher $\HIl$ masses than UNAM-KIAS, which we attribute to higher local densities probed by the latter. Furthermore, both samples show bar frequencies similar to normal spirals, indicating that the gravitational instabilities leading to bars predominantly stem from internal processes, as suggested by recent works. Our galaxies show unexpectedly high merger fractions, possibly due to sampling bias and/or the inability of the classification method to distinguish between flybys and encounters leading to coalescence. We also find higher sSFRs for asymmetric galaxies below $\mstar\sim 10^{10.3}\,\msol$, in agreement with the predictions for centrals from the \textsc{\Large eagle} simulation. We release the $\HIl$ measurements along with ancillary galaxy and halo properties for public use.     
\end{abstract}


%%%%%%%%%%%%%%%%%%%%%%%%%%%%%%%%%%%%%%%%%%%%%%%%%%

%%%%%%%%%%%%%%%%% BODY OF PAPER %%%%%%%%%%%%%%%%%%
\section{Introduction}
Neutral atomic hydrogen ($\HI$) is the predominant component of the atomic phase of the interstellar
medium (ISM), and plays a critical role in the growth and transformation of galaxies. It does so by fuelling star formation via the conversion of $\HI$
to molecular hydrogen ($\Hmol$) on dust grains, and by providing rotational support to the galaxy through its angular momentum. This means that understanding the factors that modulate $\HI$ properties is key to unravelling
the intricacies of galaxy evolution. 

It is well established that $\HI$ is readily observed as the 21-cm emission. This line was first predicted by \citet{Hulst1945} as a forbidden line arising from the spin-flip transition, but was subsequently confirmed observationally by \citet{Ewen1951}, \citet{Muller1951}, and \citet{Pawsey1951}, owing to the vast abundance of $\HI$ in our Universe. Since then, the $\HI$ line has become a cornerstone in galaxy evolution studies, providing deep insights into the structure, dynamics, and physical conditions of atomic gas in galaxies, including the Milky Way (see the review by \citealt{Dutta2022}).

We are entering an era of unprecedented observational capabilities with large optical surveys like DESI \citep{Dey2019}, Euclid \citep{Mellier2025}, and the upcoming LSST \citep{Ivezic2019}, enabling detailed mapping of billions of galaxies. These surveys will be complemented by next-generation radio observatories such as the Square Kilometre Array (SKA; \citealt{Dewdney2009}), and intensity mapping experiments such as MeerKLASS \citep{Wang2021}, CHIME \citep{Amiri2022}, and Tianlai \citep{Chen2011}. Together, these efforts promise a comprehensive, multi-wavelength characterisation of galaxies and their environments at fine spatial resolutions.

Most of the current $\HI$ observations are available as single-dish (and therefore spatially unresolved) spectra,
and are likely to remain so for many years to come, particularly for distant galaxies. 
Nevertheless, these integrated/global spectra contain valuable information about $\HI$'s spatial distribution and kinematics, and have revealed clear signs of asymmetries and disturbances in a substantial fraction ($\gtrsim 30$ per cent)
of galaxies \citep[e.g.][]{Richter1994,Haynes1998,Matthews1998,Watts2020,Watts2021}. Such asymmetries are intriguing from a theoretical standpoint because lopsided orbits are expected to dwindle within a few dynamical times due to gas dissipation and differential rotation within the disk \citep{Bournaud2005,Chakrabarti2011}. Hence, the observed incidence requires that asymmetries be induced frequently, which has motivated numerous investigations of their sources.

A host of phenomena have been proposed to explain $\HI$ line asymmetries. For the galaxies residing in dense environments, asymmetries are often attributed to ram pressure exerted by the intrahalo gas
and/or tidal interactions with other galaxies or the host halo \citep{Scott2018,Bok2019,Reynolds2020,Watts2020,Michalowski2021}. Other mechanisms have been proposed for more isolated galaxies, including gas accretion \citep{Mapelli2008,Sancisi2008}, stellar or
active galactic nuclei (AGN) feedback \citep{Manuwal2022}, and lopsided dark matter haloes \citep{Jog2009}. However, the relative contribution from these processes remains unclear. 

Disentangling the contribution of individual processes in observations is, of course, cumbersome, if not downright impossible. It should nevertheless be feasible to quantify the collective contribution of external factors by comparing against a control sample of isolated galaxies. This is contingent
on the premise that external perturbers add to the disturbances caused by \textit{in situ} phenomena, and therefore, must elevate the $\HI$ line asymmetry relative to that of an average isolated galaxy.
Several isolated $\HI$ samples have already been used for this purpose \citep{Richter1994,Haynes1998,Matthews1998,Espada2011,Bok2019}, but their isolation robustness has been revised over the years. 

The samples by \citet{Richter1994} and \citet{Matthews1998} are not strictly isolated and comprise
disc-dominated or spiral field galaxies, likely contaminated by galaxies in pairs or small groups. The first well-defined isolation criteria were proposed by \citet{Haynes1998}, who required a minimum angular and velocity separation of $1^\circ$ and $400~\kms$, respectively, for the galaxies in the Arecibo General Catalog (AGC). \citet{Espada2011} sample is based on the initial AMIGA sample, which required that, for a main galaxy with diameter $D$, there should be no galaxies with diameters $0.25D<d<4D$ located within $20\times d$ in projection. \citet{Bok2019} imposed a minimum projected separation of 500~kpc and a velocity separation of $5000~\kms$, but only focused on the galaxies covered by the Arecibo Legacy Fast ALFA survey (ALFALFA; \citealt{Haynes2018}).

Among these samples, the recent ones are clearly better suited to serve as references, and are often used as such \citep{Watts2020,Reynolds2020,Reynolds2020b,Glowacki2022,Yu2022,Zuo2022}. However, there is enough scope for improvement to warrant novel isolated $\HI$ samples, especially in light of
the available optical catalogs that offer superior completeness, isolation, and statistics. This includes the UNAM-KIAS sample \citep{Hernandez-Toledo2010} and the current AMIGA sample of isolated galaxies \citep{Argudo-Fernandez2015}, which uses a more rigorous isolation strategy than the one used to build the sample in \citet{Espada2011}.

Furthermore, recent studies have highlighted several aspects that can unduly bias the inferred asymmetry but are often neglected in the literature. \citet{Deg2020} and \citet{Watts2020} demonstrated that $\HI$ line asymmetries are markedly inflated at low values of signal-to-noise ratio ($S/N$), with the exact magnitude depending on the measure used to quantify the asymmetry; the measures capturing finer-scale variations converge at higher $S/N$ \citep{Deg2020}. The authors also 
showed that the $\HI$ line profiles need to be resolved with an ample number of velocity channels to achieve well-converged asymmetries. In addition, they found that asymmetry can vary strongly with the line of sight towards the galaxy, spanning from near-perfect symmetry to highly asymmetric, implying that an asymmetric
$\HI$ line represents intrinsic asymmetry in the galaxy's $\HI$, but a symmetric profile does not always correspond to a symmetric galaxy.

Later, \citet{Manuwal2022} analysed the $\HI$ lines of galaxies in the \textsc{\large eagle} simulation \citep{Schaye2015} and proposed a new technique for identifying profile edges based on integrated flux. This reduces the sensitivity of profile
edges to noise and yields profile widths closer to the maximum circular velocity of the underlying dark matter halo. They also showed that unresolved $\HI$ profile asymmetries are highly sensitive to observational factors, like instrumental noise, inclination, and distance to the galaxy,
and that accounting for these factors is necessary when comparing observed asymmetries with
those from simulations. They also found that differences in asymmetry are more pronounced when 
galaxies are viewed edge-on. Additionally, they found that galaxies with asymmetric $\HI$ profiles typically have lower $\HI$ fractions, higher $\Hmol$ fractions, less ordered rotation, and reside in dynamically unrelaxed dark matter halos with more substructure. This suggests that one must sample a diverse range of galaxies and environments to draw broader conclusions.

The present work aims to establish rich and reliable reference sets of isolated galactic $\HI$ sources for precise investigations of the intrinsic drivers of $\HI$ asymmetries in galaxies across a variety of environments. We use the UNAM-KIAS sample of isolated galaxies \citep{Hernandez-Toledo2010}, constructed by imposing controlled environmental conditions on Data Release 5 of the Sloan
Digital Sky Survey \citep[SDSS DR5;][]{Adelman-McCarthy2007}. We additionally quantify the $\HI$ line asymmetries in the latest version of the AMIGA sample of isolated galaxies extracted from SDSS DR10 \citep{Argudo-Fernandez2015}, one of the most well-curated and utilised reference samples to date. Our samples benefit from extensive single-dish $\HI$ data gathered from major radio observatories, such as Arecibo, Green Bank, and Nancay, and incorporate recent corrections and biases highlighted by \citet{Deg2020}, \citet{Watts2020}, and \citet{Manuwal2022} to ensure robust asymmetry measurements. To enable multivariate analyses, the samples also include inclination, stellar mass, star formation rate, optical morphology, bar presence, stellar asymmetry, and halo mass. In addition, we provide merger information from \citet{Nevin2023} based on a probabilistic assessment using linear discriminant analysis (LDA) with seven structural parameters estimated from SDSS images.

This paper serves as a detailed presentation of our catalog and is intended to provide insights that enable its proper and informed use by the community. The organization is as follows. Section~\ref{data} describes the parent isolated samples (UNAM-KIAS and AMIGA) and the various $\HI$ surveys used for retrieving the $\HI$ spectra. Section~\ref{method} details the construction of the isolated $\HI$
samples, the analysis of the $\HI$ lines therein, and the acquisition of ancillary galaxy and halo properties. In Section~\ref{results}, we discuss stellar mass ($\mstar$), inclination, $\HI$ fraction, star formation rate (SFR), morphology, bar presence, halo mass, and merger type/stage, and explore potential biases due to these properties in the asymmetries of our galaxies.  Section~\ref{previous} compares our isolated $\HI$ samples against previous samples in the literature. Finally, we present a concise summary of this work and our main conclusions, along with comments on future work, in Section~\ref{summary}.

Throughout this paper, we assume the \citealt{Aghanim2020} cosmology where necessary -- i.e. $h=0.677, \Omega_{\rm m}=0.311$, and $\Omega_{\Lambda}=0.689$. All the logarithms are base-10.

\section{Data}\label{data}
\subsection{Parent samples of isolated galaxies}
We use the UNAM-KIAS and AMIGA samples of isolated galaxies, which are minimally affected by external interactions, as our reference samples for studying the intrinsic $\HI$ line asymmetries of galaxies. In addition to studies focused on the asymmetries of isolated galaxies, these samples are well-suited for comparisons with galaxy populations in denser environments, such as groups or clusters. In the following, we briefly describe the selection criteria and highlight some strengths that make both the UNAM-KIAS and AMIGA samples appropriate for our present study, and the key differences between them. 

\subsubsection{UNAM-KIAS}\label{uk}
The UNAM-KIAS catalog \citep{Hernandez-Toledo2010} is a well-defined sample of isolated galaxies developed through a collaboration between Universidad Nacional Aut\'onoma de M\'exico (UNAM) and Korean Institute for Advanced Study (KIAS). The sample is drawn from the DR4plus release of the New York University Value-Added Galaxy Catalog (NYU-VAGC; \citealt{Blanton2005}), part of the Sloan Digital Sky Survey Data Release 5 (SDSS DR5; \citealt{Adelman-McCarthy2007}). The authors used a parent sample of 312,338 objects within a survey area of 4,464 deg$^2$ and extinction-corrected apparent magnitudes $14.5\leq m_r< 17.6$, along with an additional sample of 5,195 galaxies with $m_r<14.5$ and redshifts compiled from the literature \citep[see][]{Choi2007}. The sample was designed with strict environmental selection criteria to identify galaxies at $z\lesssim 0.3$
evolving in low-density environments, thereby minimising the influence of interactions with neighbouring systems. 
The isolation criteria were as follows:
\begin{enumerate}
    \item The magnitude gap ($\Delta m_r$) between the candidate galaxy and the neighbour is $\geq 2.5$ mag;
    \item The projected distance between the two is $\Delta d\geq 100R$, where $R$ is the seeing-corrected Petrosian
    radius of the candidate galaxy measured in the $i$-band; and
    \item The line-of-sight velocity separation $|\Delta v|\geq 1000\,\kms$.
\end{enumerate}
Additional quality controls excluded galaxies affected by poor image deblending, diffraction spikes, bright stars, or diffuse light contamination. This corresponds to a final sample of 1520 galaxies.

Compared to earlier compilations like the Catalog of Isolated Galaxies (CIG; \citealt{Karachentseva1973}), the UNAM-KIAS catalog applies more rigorous isolation thresholds. It not only recovers truly isolated galaxies in the local Universe ($v<10,000\,\kms$) but also extends the sample to higher redshifts up to $\sim 20,000\,\kms$, providing a broader basis for the statistical study of galaxy evolution in isolation. This is evident by the fact that
most of these galaxies ($\approx$ 80 per cent) exhibit morphologies ranging from Sa to Sm types \citep{Hernandez-Toledo2010} and show colours, SFRs, and $\mstar$s similar to those in central galaxies in the field \citep{Lacerna2014}. 
The galaxies also show slightly redder colours than isolated triplets, which is attributed to interaction-driven enhancement of star formation activity \citep{Hernandez-Toldedo2011}. The ellipticals in this sample are akin to those in galaxy clusters in terms of their SFRs and sizes, but present considerably greater incidence of blue and star-forming (SF) cases \citep{Lacerna2016}. 

Nonetheless, we also note that the isolation strategy adopted here does not prohibit the presence of minor/faint companions near these galaxies. As shown in the next subsection, this is important for interpreting the differences with AMIGA.

\subsubsection{AMIGA}\label{amiga}
Analysis of the interstellar Medium of Isolated Galaxies (AMIGA\footnote{\url{http://amiga.iaa.es/}}; \citealt{Verdes-Montenegro2005}) is a project focused on studying the ISM of isolated galaxies through multi-wavelength observations. It uses CIG \citep{Karachentseva1973} as a starting point and extracts a refined sample by revisiting galaxy positions, sample selection, magnitude correction, morphological characterisation, and degree of isolation. The sample has undergone two revisions since the
project's commencement. We use their latest (SIG) sample of isolated galaxies \citep{Argudo-Fernandez2015},
which was constructed using a `primary' sample containing the candidate isolated objects, and a `neighbour' sample composed of neighbours of these candidates. 

The primary sample was drawn from the main spectroscopic sample \citep{Strauss2002} of SDSS DR10 \citep{Ahn2014}, with magnitudes in the range $11\leq m_r \leq 15.7$. Only galaxies at $0.005\leq z \leq 0.080$ were considered, with the lower redshift limit set to avoid nearby sources with a dubious degree of isolation. This sample was further cleaned using the approach by \citet{Argudo-Fernandez2014}, which selects all galaxies within at least a 1 Mpc radius that are fully sampled by the SDSS DR10 photometric catalog, rejects stars misclassified as galaxies, and discards redundant identifications.

The neighbour sample was built by searching regions around the primary sample and including objects within 1 Mpc that satisfy $0.001<z<0.1$. These searches were performed in
the main SDSS spectroscopic sample and the Baryonic Oscillation Spectroscopic Survey (BOSS; \citealt{Dawson2013}). As with the primary sample, this sample was cleaned of misidentified stars and duplicate objects. 

The parent sample constructed in this way comprised of 33081 primary galaxies with 1607947 neighbours. The isolated objects among the primary galaxies were identified by requiring that there is no neighbour with $\Delta d\leq 1$ Mpc and $|\Delta v|\leq 500\,\kms$. Since the SDSS spectroscopic sample is complete over $14.5\lesssim m_r\lesssim 17.7$ \citep{Strauss2002}, only those primary galaxy fields were considered in which at least 80 per cent of the extended neighbours (within 1 Mpc) have measured spectroscopic redshifts. This sample was further refined by excluding the objects affected by fibre collisions and mergers, resulting in a final sample of 3702 galaxies. Out of these, 449 ($\approx 12$ per cent) are in common with UNAM-KIAS (table 4, \citealt{Argudo-Fernandez2015}).

Similar to UNAM-KIAS, the AMIGA sample is dominated by spirals \citep{Fernandez2012,Buta2019} and presents elliptical sizes consistent with
those in clusters \citep{Fernandez2013}. However, massive spirals in AMIGA
seem to be larger than those in denser environments \citep{Fernandez2013}. In addition,
they show a higher incidence of optical AGN and sSFRs than typical galaxies,
where the former is linked to cold gas accretion \citep{Argudo-Fernandez2016}. They also seem to exhibit higher $\HI$ masses than galaxies in clusters \citep{Jones2018,Calette+2018}.

Note that, even though UNAM-KIAS appears more restrictive due to lower statistics, the galaxies in AMIGA are evidently more isolated. This has already been demonstrated in \citet{Argudo-Fernandez2015}, which shows that galaxies in UNAM-KIAS have neighbours separated by $\Delta d\leq 1$ Mpc and $|\Delta v|\leq 500\,\kms$ (see fig. 5 in that paper). We attribute these differences to two main factors. Firstly, UNAM-KIAS requires that there be no companions brighter than $\Delta m_r=2.5$ whereas AMIGA considers the absence of \textit{all} the galaxies covered by the survey. Secondly, the distance cut in UNAM-KIAS is 100 times the (Petrosian) size of the main galaxy. These sizes span $\approx 2-5$~kpc for the $\mstar$ range explored here \citep{Shen2003}, resulting in distance cuts of $\lesssim 0.5$~Mpc, shorter
than the fixed threshold of 1 Mpc used for AMIGA. Hence, the better statistics in AMIGA is an outcome of using SDSS DR10, which offers a larger footprint and spectroscopic coverage than the earlier data release used for building UNAM-KIAS.

\subsection{\HIn line profiles}\label{hisamples}
We source the $\HI$ profiles of the isolated galaxies from five extensive 21-cm emission line surveys. These are
described as follows.
\subsubsection{$\HIi$-MaNGA}
$\HI$-MaNGA \citep{Masters2019,Stark2021} is a 21-cm follow-up program for galaxies in the SDSS-IV MaNGA survey \citep{Bundy2015,Yan2016,Blanton2017} conducted through the $L$-band receiver on the Green Bank Telescope (GBT; \citealt{gbt}). All galaxies with $z\lesssim 0.05$ without $\HI$ data were targeted regardless of their stellar mass, morphology, or colour. This corresponds to stellar masses in the range $10^{8.7}\lesssim \mstar/\msol \lesssim 10^{11.5}$. The spectra exhibit the {\it raw} (or true) spectral
resolution of $v_{\rm res}\sim 1.2\,\kms$, and were cleaned of features due to strong radio frequency interference (RFI), and boxcar and Hanning smoothed to an \textit{effective} resolution of $v_{\rm eff}=10\,\kms$. Leftover baseline variations were removed using a polynomial fit. We use the fourth data release\footnote{\url{https://greenbankobservatory.org/portal/gbt/gbt-legacy-archive/hi-manga-data/}} \citep{Stark2021}, which also includes Arecibo observations for MaNGA galaxies ($\approx 30$ per cent) that overlap with ALFALFA (described in Section~\ref{alfalfa}). The 
total number of galaxies in this data set is 3669.

\subsubsection{NIBLES}
NIBLES\footnote{\url{http://cdsarc.u-strasbg.fr/viz-bin/cat?J/A+A/595/A118}} (Nancay Interstellar Baryon Legacy Extragalactic Survey; \citealt{vanDriel2016}) is an unbiased $\HI$ survey of galaxies in the local Universe using the 100-m class Nancay Radio Telescope. The survey targeted 2850 galaxies at $0.03<z<0.04$ in SDSS DR5 \citep{Adelman-McCarthy2007} selected to have a uniform stellar mass distribution within $10^6\lesssim \mstar/\msol \lesssim 10^{12}$. The 21-cm spectra were mitigated of RFI, fit with polynomial baselines, and smoothed to $v_{\rm eff}=18\,\kms$ ($v_{\rm res}\sim 2.6\,\kms$).

\subsubsection{KLUN}
Kinematics of the Local Universe (KLUN) is an all-sky $\HI$ survey of spiral galaxies at $z\lesssim 0.03$ compiled from the Digitized Sky Survey (DSS; \citealt{Paturel2000}), the DEep Near Infrared Survey (DENIS; \citealt{Paturel2005}), and the 2 Micron All Sky Survey (2MASS;\,\,\citealt{Jarrett2000}). The observations were conducted using the Nancay telescope over more than $20$ years \citep{Theureau1998,Paturel2003,Theureau2005,Theureau2007,Theureau2017}. The spectra were processed in a manner similar to NIBLES \citep[see][]{Theureau2005}, but smoothed to $v_{\rm eff}=10\,\kms$ ($v_{\rm res}\sim 2.6\,\kms$). For this study, we use the three public spectral data releases\footnote{\url{https://cdsarc.cds.unistra.fr/viz-bin/qcat?J/A+A/430/373}}$^,$\footnote{\url{https://cdsarc.cds.unistra.fr/viz-bin/qcat?J/A+A/465/71}}$^,$\footnote{\url{http://cdsarc.u-strasbg.fr/viz-bin/qcat?J/A+A/599/A104}} from the survey \citep{Theureau2005,Theureau2007,Theureau2017},
corresponding to a total of 2173 galaxies.

\subsubsection{ALFALFA}\label{alfalfa}
Arecibo Legacy Fast ALFA (ALFALFA; \citealt{Haynes2018}) is a blind (untargeted) $\HI$ survey that was conducted via the Arecibo L-band Feed Array (ALFA) over $0.007 \lesssim z \lesssim 0.06$. The survey was designed to sample the $\HI$ mass function down to the faint end within a volume of $\approx 100^3\,{\rm Mpc}^3$. 
As such, it is regarded as one of the most complete $\HI$ databases for the local Universe.
We use the final data release\footnote{\url{https://cdsarc.cds.unistra.fr/cgi-bin/cat/J/ApJ/861/49}} which includes $\sim 31500$ sources, whose spectra were checked for RFI, baseline subtracted, and smoothed to $v_{\rm eff}=10\,\kms$ ($v_{\rm res}\sim 5.5\,\kms$).

\subsubsection{xGASS}
xGASS refers to the extended GALEX Arecibo SDSS Survey \citep{Catinella2018}, a gas fraction-limited $\HI$ survey using ALFA of $\sim 1200$ galaxies at $0.01<z<0.05$ that lie at the intersection of SDSS DR7 \citep{Abazajian2009}, the GALEX Medium Imaging Survey \citep{Martin2005}, and ALFALFA. The galaxies were selected to ensure even sampling across $10^9<\mstar/\msol<11.5$ based on the stellar masses in the MPA-JHU catalog\footnote{\url{http://home.strw.leidenuniv.nl/~jarle/SDSS/}},
and were observed until detected or a certain gas fraction is reached. For the latter, it was ensured that $\log(M_{\HIs}/\mstar)=8$ for $\log(\mstar/\msol)<9.7$, and $M_{\HIs}/\mstar>0.02$ for $\log(\mstar/\msol)>9.7$. This ensures that the gas fraction limit spans $[0.02,1]$. The spectra\footnote{\url{https://cdsarc.cds.unistra.fr/viz-bin/cat/J/MNRAS/476/875}} were bandpass-subtracted, RFI-excised, and smoothed $v_{\rm eff}=5-15\,\kms$ ($v_{\rm res}\sim 1.4\,\kms$).

\section{Methods}\label{method}
\subsection{Constructing the isolated \HIn samples}
The primary goal of the present work is to conduct a careful search across five public $\HI$ archives (Section~\ref{cross}) and identify, among the parent samples, the galaxies whose $\HI$ spectra are obtained with proper beam coverage (Section~\ref{beam}), have high signal-to-noise ratio, and are resolved with sufficient number of velocity channels (Section~\ref{sbyn}).

\subsubsection{Cross-matching with the $\HIi$ archives}\label{cross}
We search for the $\HI$ profiles of galaxies in each of the two parent isolated samples. Since a galaxy could appear in multiple $\HI$ surveys, we perform the cross-matching to ensure that the included spectrum is the one observed with the largest possible beam size.
This is done to maximise the coverage of large-scale perturbations 
that are common in extended $\HI$ disks \citep{Zheng2022} and can exist in isolated galaxies \citep[e.g.][]{Sengupta2012,Scott2014}. The $\HI$ surveys used in this study are based on these telescopes ordered by decreasing beam (FWHM) dimensions: GBT ($9.17^\circ\times9.14^\circ$), Nancay ($4^\circ\times22^\circ$), and Arecibo ($3.1^\circ\times3.5^\circ$). Usually only one of these was used in a given survey, except for $\HI$-MaNGA, which includes observations
from both GBT and Arecibo. 

We begin our search by identifying the $\HI$-MaNGA galaxies with GBT data that fall within GBT's beam width and have $\Delta v<200\,\kms$. If a match is found, we further check if there are multiple galaxies within 1.5 times the beam width and $\Delta v<200\,\kms$ in the NASA-Sloan Atlas\footnote{\url{https://www.nsatlas.org/}}. The
second check addresses the possibility of untargeted galaxies with $\HI$ lying within the beam. We accept the spectral match only if a single match is found in both of these checks. This ensures a unique match and diminishes the chance of contamination from $\HI$ in another galaxy
in the proximity (or `$\HI$ confusion'). For the isolated galaxies without matches after this search, we perform a similar search in NIBLES using Nancay's beam width, followed by a search in KLUN. For the remaining isolated galaxies, we formulate the search criteria based on Arecibo's beam size and search in ALFALFA, the $\HI$-MaNGA subsample with Arecibo data, and xGASS. The exercise yields 426 matches for UNAM-KIAS and 903 for AMIGA.

\subsubsection{Beam coverage}\label{beam}
Selecting the match corresponding to the largest telescope ensures the widest possible beam coverage, but it does not guarantee that the beam covers all the $\HI$ in the galaxy. This is strictly 
valid only if the galaxy is located beyond a threshold distance that depends on the telescope. In this work, we have devised a strategy to determine whether the matched $\HI$ spectrum indeed represents all the $\HI$. 

Resolved 21-cm observations of galaxies in the local Universe have shown that the $\HI$ size, defined as the radius where the $\HI$ surface density drops
to $1\,\msol{\rm pc}^{-2}$, exhibits a tight correlation with the $\HI$ mass encapsulating the full $\HI$ content of the galaxy \citep{Broeils1997,Swaters2002,Begum2008,Wang2013,Lelli2016,Wang2016}. This means that, for a galaxy with near-complete beam coverage, the linear beam size at its distance ($D_{\rm beam}$) must be greater than its $\HI$ diameter ($D_{\HIs}$). We explore the possibility of using this relation to assess beam coverage for our galaxies.

\citet{Wang2016} showed that $\Sigma_{\HIs}$ profiles are roughly self-similar once they are normalised by the $\HI$ radius ($R_{\HIs}$), meaning that galaxies have similar fractions of their $\HI$ masses located within a circular aperture of radius $r=fR_{\HIs}$,
as shown explicitly by the authors. We use the results in \citet{Wang2016} to determine the mean $\HI$ mass coverages expected for beams with $D_{\rm beam}=fD_{\HIs}$ for different values of $f$, and the $\HI$ diameters ($D_{\HIs,\,\rm pred}$) corresponding to these masses predicted by the mean $\HI$ size--mass relation. This allows us to obtain a relationship between $\HI$ coverage by the beam and $D_{\rm beam}/D_{\HIs,\,\rm pred}$. We find that $D_{\rm beam}/D_{\HIs,\,\rm pred}$ increases for higher $f$ and that $D_{\rm beam}\lesssim D_{\HIs,\,\rm pred}$ for $f\lesssim 0.9$, corresponding to $\HI$ mass recovery of about 80 per cent or less.

Motivated by these results, for our sample, we adopt the $M_{\HIs}$ from the catalog of the $\HI$ survey that the galaxy's 21-cm spectrum was sourced from, use the mean \citet{Wang2016} size--mass relation to compute the $D_{\HIs,\rm pred}$ for this mass, and exclude the match if $D_{\rm beam}/D_{\HIs,\rm pred}<1$. We find that virtually all the galaxies satisfying this cut have $D_{\rm beam}\gtrsim 1.5D_{\HIs,\,\rm pred}$, implying that more than 90 per cent of the total $\HI$ was observed\footnote{Beam coverages may have been slightly over- or under-estimated because of the intrinsic scatter in the $\HI$ mass coverages and the size--mass relation in \citet{Wang2016}. However, we think that this does not plague our sample given that $D_{\rm beam}/D_{\HIs,\rm pred}$ values are sufficiently high to ensure wide coverages despite the aforementioned uncertainties.}. The cut reduces our UNAM-KIAS matched sample to 408, and the AMIGA sample to 899.

\subsubsection{Profile characterisation and signal-to-noise threshold}\label{sbyn}
The analysis of an $\HI$ line is usually carried out only within a pre-determined region of the profile. This is because the profile outskirts are dominated by noise and are unreliable for measurements. The approach by \citet{Manuwal2022} involves deriving
the profile edges using the best-fit Busy function \citep{Westmeier2014} for the spectrum. This is done by calculating the flux-weighted velocity of the fit, splitting it into two sides about this velocity, and integrating towards the lower and higher velocity sides until we reach 95 per cent of the flux on each side. This results in edges $v_l$ and $v_h$ for the lower and higher velocity sides. With these edges determined, one can compute the asymmetries with the channels bounded within the edges in the \textit{true} spectrum. 

The lopsidedness, $A_l$, measures the difference between the integrated fluxes on the low-velocity ($F_l$) and high-velocity ($F_h$) sides about the systemic velocity ($v_{\rm sys}$, i.e. the mid point of the profile edges), and is calculated as 
\begin{equation}
  A_l = \frac{|F_l - F_ h|}{F_l + F_h}.
  \label{lop}
\end{equation}
This measure captures global asymmetry in the $\HI$ distribution and is a reformulation of the `flux ratio' ($A_{fr}$) measure often used in the literature \citep[e.g.][]{Haynes1998,Espada2011,Reynolds2020,Reynolds2020b,Watts2020}, defined as
\begin{equation} \label{eq:Afr}
    A_{fr} = 
        \begin{cases}
            \frac{F_l}{F_h} & F_l \geq F_h\\
           \frac{F_h}{F_l} & F_l < F_h.
        \end{cases}
\end{equation}
However, unlike $A_{fr}$, $A_l$'s dynamical range is by construction restricted to $0 \leq A_l \leq 1$, with $A_l=0$ implying perfect symmetry.

As a rule of thumb, $A_l\gtrsim 0.1$ indicates strong lopsidedness that is apparent as a visibly disturbed $\HI$ disk \citep{Manuwal2022}. It is important to note, however, that the converse may not always be true -- that is, a genuinely asymmetric galaxy may not always exhibit a high $A_l$. Similarly, it follows that not every galaxy with a low $A_l$ has a symmetric $\HI$ distribution. This is because the global $\HI$ profile captures only line-of-sight information and is therefore sensitive to the viewing angle (see \citealt{Deg2020} for a discussion). Although this precludes deriving robust conclusions 
for individual galaxies with symmetric $\HI$ line profiles, it is nevertheless possible to infer systematic differences between two populations of randomly-oriented galaxies. (We come back to this when
we discuss the importance of inclination in Section~\ref{inc}).

In addition to $A_l$, \citet{Manuwal2022} quantified $\HI$ line asymmetry using the `velocity offset' ($A_{vo}$) and the `normalised residual' ($A_{nr}$), where the former captures the deviation of $v_{\rm sys}$ from the velocity corresponding to $A_l=0$, and the latter gauges
the flux offsets between pairs of equally-spaced velocity bins about $v_{\rm sys}$. These measures are correlated but with some scatter, because they capture different information about profile asymmetry \citep{Manuwal2022}. It is therefore preferable to use all three. However, this may not always be feasible as some of these measures are more susceptible to noise than others. Our tests on model $\HI$ profiles with varying degrees of intrinsic (noiseless) asymmetry reveal that $A_l$ generally requires the least $S/N$ to converge within a given tolerance (Appendix~\ref{conv}). Given this, we use only $A_l$ in this work to maximise the statistics of cases with robust asymmetries.

We determine the $S/N$ in
accordance with \citet{Saintonge2007}:
\begin{equation}
S/N = \frac{1}{\sigma_{\rm rms}W_{95}}\sqrt{\frac{W_{95}}{2v_{\rm eff}}} \sum_{v=v_l}^{v_h}F(v)v_{\rm res},    
\end{equation}
where $W_{95}=v_h-v_l$ is the linewidth, $\sigma_{\rm rms}$ is the RMS noise, and $F(v)$ is the flux level at velocity $v$. We choose $S/N>10$ as the threshold, which is sufficient to achieve convergence within $\approx 20$ per cent for cases with minor asymmetries ($A_l\sim 0.05$). This convergence behaviour is fairly robust against variations in $v_{\rm res}$ and $v_{\rm eff}$, at least for the range spanned by the spectra considered in this work (Appendix~\ref{conv}). Profiles with even lower asymmetries require exceedingly high $S/N$ but are not predicted to dominate the statistics \citep{Watts2020,Glowacki2022,Manuwal2022}, and we do not expect them to contaminate our results in any notable way. This results in 217 galaxies from UNAM-KIAS and 398 galaxies from AMIGA.

\subsubsection{Velocity channels and magnitude limit}\label{chanandmag}
Asymmetry measurements can also be affected by systematic biases if the profile is not resolved with a sufficient number of velocity channels. We account for this by following the recommendation of \citet{Deg2020} and requiring that our profiles be resolved with $\geq 20$ velocity channels. The resulting $\HI$ samples have 154 UNAM-KIAS and 348 AMIGA galaxies.

Furthermore, we note that
AMIGA, by construction, reaches fainter magnitudes than UNAM-KIAS. This effectively introduces a bias towards smaller stellar masses in the former. Therefore, for the sake of homogeneous completeness in our samples, we include only galaxies with $m_r<15.2$. We use the Petrosian magnitudes provided in the UNAM-KIAS catalog and the magnitudes obtained for AMIGA via cross-matching with the SDSS DR10 database using a $1''$ search radius\footnote{This is the search window used for all the cross-matches in this paper, unless otherwise stated}. Three AMIGA galaxies do not match and are excluded. This yields final samples comprising 153 objects from UNAM-KIAS and 236 from AMIGA, which are the samples we work with throughout the paper.

Recall that UNAM-KIAS and AMIGA have considerable overlap \citep[see][]{Argudo-Fernandez2015}, meaning that one needs to keep a note of such cases when comparing these samples. We find the redundant objects by searching for galaxies in these catalogs with the same matched $\HI$ spectrum, which reveals 59 cases. We discuss the impact of these galaxies on the differences observed between the two samples for various properties in Section~\ref{results}.

\subsection{Acquisition of galaxy and halo properties}
\subsubsection{$\HIi$ mass, stellar mass, and star formation rate}
We evaluate the $\HI$ mass ($M_{\HIs}$) using the luminosity distance of the galaxy [$D(z)$] and the integrated $\HI$ flux within the profile edges ($F_{\HIs}$) as
\begin{equation}\label{himass}
\frac{M_{\HIs}}{\msol} = \frac{2.36\times10^5}{(1+z)^2}\left(\frac{D}{{\rm Mpc}}\right)^2\left(\frac{F_{\HIs}}{{\rm Jy\,\kms}}\right).    
\end{equation}

The approach for $\mstar$ and SFR is based on the consideration that the constraints on these quantities improve with the wavelength's dynamic range in the spectral energy distribution (SED). We adopt the GSWLC-X2\footnote{\url{https://salims.pages.iu.edu/gswlc/}} catalog \citep{Salim2016,Salim2018} as our standard, where galaxy properties are derived from spectral bands spanning far-UV to mid-IR (GALEX+SDSS+WISE photometry). The SED fitting employs the \citet{Bruzual2003} stellar population synthesis model and the \textsc{\large cigale} SED fitting code \citep{cigale}, assuming an energy balance between stellar emission and its absorption and re-emission in the IR by dust [akin to other extensively-used codes like \textsc{\large prospect} \citep{prospect} and \textsc{\large magphys} \citep{magphys}].

We begin by searching for a counterpart in GSWLC-X2 and adopt the values from therein if available. Otherwise, we use values that are considered close GSWLC-X2 equivalents. For this, we first check whether the corresponding $\HI$ spectrum is from ALFALFA, and if so, opt for the values in the ALFALFA-SDSS catalog \citep{Durbala2020}. This catalog contains estimates from various methods, and we use the GSWLC-X2 values if provided. If not, we use $\mstar$ derived by \citet{Durbala2020} via the \citet{Taylor2011} method and SDSS photometry, and convert it to the GSWLC-X2 equivalent by applying the correction presented in \cite{Durbala2020}:
\begin{equation}
\log \mstar = 1.052 \log M_{\star,\rm Taylor}-0.369.     
\end{equation}
Likewise, for the SFR, we get the GSWLC-X2 equivalent using the NUV+IR value in ALFALFA-SDSS if available as
\begin{equation}
\log {\rm SFR} = \log {\rm SFR_{\rm NUV+IR}+0.09},    
\end{equation}
and use the MIR (22$\mu$m) value otherwise:
\begin{equation}
\log {\rm SFR} = \log {\rm SFR_{\rm 22}-0.09}.    
\end{equation}

If we are unable to find any match, or the spectrum is not from ALFALFA, we perform a cross-match against the \citet{Chang2015} catalog. The catalog provides the physical properties derived through SED fitting of SDSS+WISE photometry using \textsc{\large magphys}. If a match is found, we increase the $\mstar$ by 0.1 dex to account for the typical offset for SF galaxies relative to GSWLC-X2 \citep{Salim2016}, whereas the SFR is adopted as is. If no match has been found yet, we finally conduct a similar search within 1.5$''$ \citep[as used by][for example]{Li2023} in the DESI value-added catalog \citep{Siudek2024}, which contains estimates based on DESI+WISE photometry and \textsc{\large cigale} fits. If we find a match, the $\mstar$ is converted to 
GSWLC-X2 equivalent as per the relation in fig. C2 of \citet{Siudek2024}, i.e.
\begin{equation}
\log \mstar = 0.980 \log M_{\star,\rm DESI}+0.471,     
\end{equation}
and take the SFR as is. This way, we are able to obtain stellar masses and SFRs for $\approx 94$ per cent of our UNAM-KIAS sample and $\approx 99$ per cent of our AMIGA sample.

\subsubsection{Inclination}\label{inc}
Ideally, one should use a resolved $\HI$ map of the galaxy to infer its inclination, as the $\HI$ disk may not be perfectly aligned with the stellar component. Nevertheless, since $\HI$ is a cold component of the ISM, the two are not expected to be strongly misaligned in most cases and it is a common practice to use the inclination derived from stars \citep[e.g.][]{Verheijen2001,Bok2019,Yu2022}. Our samples specifically
deal with isolated galaxies, which have even smaller likelihoods of such misalignments. We therefore use the inclinations derived from optical images in this work.

We estimate the inclinations of our galaxies using their $r$-band images, as the band offers a good balance between signal-to-noise and dust attenuation. First, we cross-match against the \citet{Simard2011} catalog and use the disk inclination for the bulge+disk model with a de Vaucouleur's bulge if it is preferred over a free-Sersic bulge with a probability greater than 50 per cent, and the latter otherwise. If no match is found, we assume the minor-to-major axis ratio ($b/a$) of the disk from the best-fit model in the \citet{Meert2015} catalog and convert it to inclination as
\begin{equation}
i = {\rm sin}^{-1}\left(\sqrt{\frac{1-(b/a)^2}{1-q^2}}\right),
\end{equation}
where $q=0.2$ is the assumed intrinsic thickness \citep[e.g.][]{Catinella2012a,Stark2021}. The inclination is set to $90^\circ$ when $b/a<q$. If no match is found, the inclination is determined using the axis ratio for the two-dimensional Sersic model in the NSA catalog. Finally, if there is still no match at this stage, we opt for the inclination corresponding to the axis ratio for the model
with the greater log-likelihood in the SDSS DR10 photometric catalog (see \citealt{Stoughton2002} for details). This covers our full AMIGA sample and $\approx 80$ per cent of UNAM-KIAS.

We do not exclude galaxies at low inclinations because we prefer to leave the choice of the appropriate threshold to the user's discretion for a few practical reasons. For one, the definition of `low' is arbitrary, and one may not always have enough statistics to conduct science with highly inclined populations. Moreover, it may not even be necessary to impose a threshold if the samples are strongly disparate in their asymmetry levels, as has been previously demonstrated for centrals and satellites, for example \citep{Watts2020,Watts2020b,Manuwal2022}.

\subsubsection{Stellar morphology and bar presence}\label{morph}
For the UNAM-KIAS galaxies, we adopt the morphological classifications presented by \citet{Hernandez-Toledo2010}. This scheme is based on visual inspection of SDSS DR5 images, combined with post-processing techniques to enhance morphological structures. The same catalog also provides information on stellar bars, which we incorporate into our analysis. We obtain these properties in approximately 93 per cent of our sample.

For the morphologies of AMIGA galaxies, we cross-match with the MaNGA Visual Morphology catalog \citep{Vazquez-Mata2025}, which reports visually determined Hubble morphologies from a mosaic of images obtained from SDSS and the DESI Legacy Imaging Surveys \citep{Dey2019}, along with post-processed versions that highlight morphological features. In addition, we used machine-learning-based morphological classifications of \citet{DominguezSanchez2022}. For bars, we perform our own classifications via the approach described in \citet{Hernandez-Toledo2010}. 

\subsubsection{Stellar asymmetry}\label{stelasym}
We adopted the two-dimensional galaxy asymmetries in the $r$-band from \citet{Nevin2023}, which are derived from the full SDSS DR7 dataset. This includes both the `Asymmetry' ($A$) and `Shape Asymmetry' ($A_S$) parameters described in the following.

The first one is computed as
\begin{equation}
    A = \sum_{ij} \frac{|I(i,j)-I_{180}(i,j)|}{|I (i,j) |} - \sum_{ij} \frac{|B(i,j)-B_{180}(i,j)|}{|I (i,j)|}
\end{equation}
where the summation is over all pixels ($i$,$j$), $I(i,j)$ is the default pixel value, $I_{180}(i,j)$ is the value after a 180$^{\circ}$ rotation about the center, $B(i,j)$ is the pixel value in the background image, and $B_{180}$ is the background level after the rotation. The center is taken as the location that minimizes $A$, and the calculation is only performed over the galaxy/background image enclosed within 1.5 times the Petrosian radius $r_{p}$. 

The analysis for $A_s$ is identical to that for $A$, but in this case the measurement is carried out on binary detection masks rather than on the images themselves \citep{Pawlik2016}. This assigns equal weight to all the pixels and mitigates the intrinsic bias in $A$ towards brighter pixels. As a result, it is relatively oblivious to asymmetries associated with bright galaxy features, such as multiple nuclei, and is more sensitive to diffuse light in the outskirts. However, $A$ can be better at
capturing asymmetries due to features such as spiral arms.

We are able to obtain these parameters for $\approx 78$ per cent and $\approx 68$ per cent of our UNAM-KIAS and AMIGA samples, respectively.

\subsubsection{Merger type and stage}\label{merger}
To identify galaxies involved in mergers, we use the machine learning-based classification developed by \citet{Nevin2019, Nevin2023}. The classification relies on a linear discriminant analysis (LDA) of seven imaging parameters that capture key features of the spatial structure of the galaxy: {\it Gini}, $M_{20}$, Concentration ($C$), Asymmetry ($A$), Smoothness ($S$), S\'ersic index ($n$), and shape asymmetry ($A_s$) measured on mock Sloan Digital Sky Survey (SDSS) $r$-band images from hydrodynamical simulations of isolated and merging galaxies. 

Among these $A$ and $A_s$ have already been described in Section~\ref{stelasym}, and the S\'ersic index is self-explanatory. $Gini$, $M_{20}$, and $C$ measure light concentration, but in different ways: $Gini$ uses pixels' ranks based on their brightness, $M_{20}$ uses the ranks and distances from the galactic center, and $C$ uses the radii containing 80 and 20 per cent of the integrated flux. $S$ is the fraction of light within clumps. [For further details, see \citet{Nevin2019,Nevin2021}].

The LDA framework uses a linear combination of these morphological features to define a hyperplane in a multidimensional parameter space. The galaxy's position relative to this hyperplane is used to compute the probability of its association with a particular merger class. This technique provides robust merger classifications across a wide range of gas fractions, mass ratios, orientations, and merger stages. The method was applied to the SDSS DR16 $r$-band images, assigning a probability for each galaxy to be in a major or minor merger (major merger if the mass ratio is $1:3$ or greater, and minor merger for smaller mass ratios) and further sub-merger stages, pre- (early, late) and post-coalescence merger (see \citealt{Nevin2019,Nevin2023} for more details). Here, `pre-merger' refers to the first pericentric passage, the first apocenter, or the final approach. The `post-merger' stage is any time within 1 Gyr of the last merger, where the timescale is inspired by the IllustrisTNG simulation \citep{Pillepich2018,Weinberger2018}, which shows disturbed morphologies for up to 2.5 Gyr \citep{Bickley2021}.

The imaging parameters are based on surface brightness maps, and do not strictly capture the galaxy's three-dimensional features. However, combining them seems to compensate for this limitation. This also facilitates the identification over a longer timescale post-/pre-merger than any individual metric. Although the training data set did not explore the full morphological range, the technique works well in practice for a diverse range of morphologies \citep[e.g.][]{Nevin2023}.

As a reference, the discriminant for major and minor mergers, as set by \citet{Nevin2023} is given by:

\begin{equation}
\begin{split}
{\rm LD_{\rm major}} = 13.9A_s -8CA_s - 5.4AA_S+5.1A + 4.8c-2.9GiniA_s\\
+0.6M_{20}A+ 0.4M_{20}n + 0.4Gini - 0.6\,,
\end{split}
\end{equation}
%and the minor merger classification is done using
\begin{equation}
\begin{split}
{\rm LD_{\rm minor}} = -10.4CA_s+8.8CA-7.8GiniS -7.8A+6.6A_s\\
+6.5GiniM_{20}-6M_{20}S-5.7M+{20}A_S+4.9S\\
-4.4M_{20}+3.7GiniC-2.9Sn-nA_s-0.2AS\\
-0.7.
\end{split}
\end{equation}
%We use these to determine probability of the merger,
Then the probability is estimated by
\begin{equation}\label{pmerg}
p_x = \frac{1}{1+e^{-{\rm LD_x}}},    
\end{equation}
where LD$_x$ is the pertinent linear discriminant. 

To avoid any bias in the prediction due to the merger fraction used as a prior in the method, \citet{Nevin2023} adopt a Bayesian marginalisation approach to calculate the probability across a range of merger fractions, resulting in 46 probabilities for each galaxy. From the resulting posterior distributions, they derived the 16th, 50th, and 84th percentiles for each galaxy, and these are the probabilities used to determine whether a galaxy is part of a merger and then distinguish between each merger stage. 

For this work, we have adopted the recipe proposed by \citet{Nevin2023} and \citet{Comerford2024} to identify mergers and determine whether they are in a pre- or post-merger stage. We consider the 50th percentile of the probability distribution for each individual galaxy and use a probability threshold of 0.9 to confidently distinguish between the classifications. A galaxy merger is considered only if the probability of a major or minor merger is $>0.9$. If both conditions are $<0.9$, then it is classified as a non-merger. Major mergers are identified as cases where the probability of a major merger is $>0.9$ and greater than the probability of a minor merger. A similar condition applies for minor merger, changing the major probability to minor and minor to major. The pre-merger stage for any merger type implies that the pre-merger probability is $>0.9$ and exceeds the post-merger probability. The post-merger stage is defined likewise. The median probabilities for both merger types and merger stages have been provided in the released catalog. We have this information for $\approx 68$ per cent of our UNAM-KIAS and $\approx 52$ per cent
of our AMIGA sample.

\subsubsection{Group membership and halo mass}\label{group}
Galaxy group catalogs are usually constructed for large surveys by iteratively assigning membership based on halo properties such as velocity dispersion and virial radius, and assuming that these properties scale with halo mass as in a $\Lambda$CDM cosmology \citep[e.g.][]{Yang2005,Yang2007,Yang2012,Lim2017,Rodriguez2020}. This outperforms traditional methods such as the friends-of-friends algorithm. In these methods, the halo mass is derived via abundance matching, where a stellar quantity (luminosity or mass) is rank-ordered, and the halo mass is taken as that of the halo with the same rank based on a theoretical mass function.

These methods work well for identifying group members and inferring the average relation with halo mass, but they introduce biases in the halo masses of galaxy populations lying at the extremes of the stellar--halo mass plane. For instance, \citet{Zhao2025} showed that abundance matching using group luminosity/$\mstar$ \citep[e.g.][]{Yang2012} introduces negative and positive biases in the halo masses of quiescent and star-forming galaxies, respectively, which artificially washes out the offset that is otherwise apparent with direct halo mass measurements from weak lensing \citep{Mandelbaum2016,Bilicki2021,Zhang2022,Zhang2024}. Newer catalogs use other quantities such as the central's stellar mass \citep{Lim2017} along with a correction for the assembly history \citep[e.g.][]{Lu2016}. While this improves the overall accuracy, it may not mitigate the aforementioned biases, particularly for haloes with a single member.

In this work, we use the group catalog for SDSS DR7 built by \citet{Tinker2021} using a novel halo-based group finder inspired by \citet{Yang2005}. Unlike prior works, their group finder includes an SFR-dependence for halo mass at fixed total group luminosity by introducing separate weights for star-forming and quiescent central galaxies during rank-ordering that are self-calibrated using two-point statistics and total satellite luminosity. This enables the group finder to improve the central/satellite classification and reproduce the observed stellar-to-halo mass relations for blue/star-forming and red/quiescent galaxies. 

We cross-match against this catalog and obtain the halo mass, central status, and number of satellites in the group for $\approx 73$ per cent of the UNAM-KIAS and $\approx 92$ per cent of the AMIGA sample. Among these galaxies, all the UNAM-KIAS objects and $\approx 96$ per cent of the AMIGA objects are isolated according to the group finder. However, even the $\approx 4$ per cent non-isolated objects in AMIGA are centrals and generally have only one satellite. This is strong evidence for the isolated nature of our galaxies.

\section{Galaxy/halo properties and their connections to $\HIlb$ line asymmetry}\label{results}

It is clear from Sections~\ref{uk} and \ref{amiga} that UNAM-KIAS and AMIGA differ in the isolation criteria adopted and may not contain similar galaxy populations. Additionally, $\HI$ line asymmetry could correlate with certain galaxy and/or halo properties within each sample, in which case a proper use of our samples for environmental studies requires awareness of such biases or their absence. In this section, we explore various properties of our galaxies and their haloes, and their trends with $A_l$, to provide insights that will not only guide future analyses of our sample but also reveal important information about the nature of isolated galaxies and the origins of asymmetries in their $\HI$ disks. This is not meant to be an exhaustive list and additional analyses may be required by the user for explorations with quantities that have not been explored directly in this section.

Briefly, our analysis reveals that there are certain trends with $\HI$ line asymmetry that are qualitatively consistent between UNAM-KIAS and AMIGA. Most of these are negative results, demonstrating $A_l$ to be generally independent of galaxy properties within our samples. This includes stellar mass, stellar (shape) asymmetry, stellar morphology, bar presence, $\HI$ fraction, and recent merger history. The only positive result presented by both samples is the correlation between $A_l$ and sSFR at lower masses. On the other hand, there are some trends that are specific to one of the two samples: $A_l$ correlates positively with galaxy inclination for UNAM-KIAS, and with rotational stellar asymmetry for AMIGA. The following text provides further details and discussion on these and some additional results.

\subsection{Stellar mass}\label{asymandmstar}
A strong dependence of asymmetry on stellar mass suggests that a trend observed
between asymmetry and a physical characteristic may actually be driven by mass bias instead of the characteristic. In addition, the dependence of asymmetry on mass is in itself an important question worth exploring from a physical standpoint. This section is devoted to an exploration of these possibilities.

\begin{figure}
    \centering
    \includegraphics[width=1\linewidth]{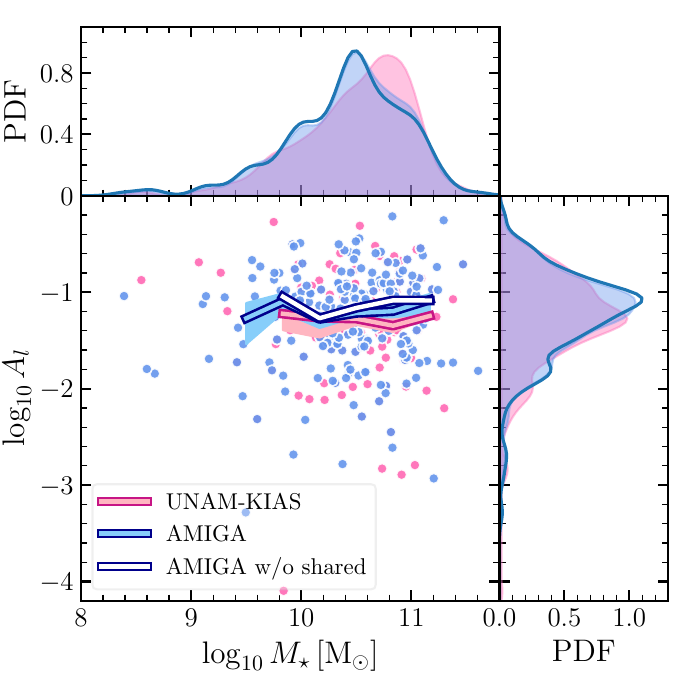}
    \caption{The relationships between $\HI$ lopsidedness and stellar mass for the isolated galaxies from UNAM-KIAS and AMIGA, shown in pink and blue, respectively. The curves are the medians and are only shown for mass bins with at least 10 points. The shaded regions show the bootstrapping errors on the medians. The open blue curve is for the AMIGA subsample without the galaxies in common with UNAM-KIAS (see the text). The wide panel at the top shows the KDE-smoothed probability distribution functions (PDFs)
    for $\mstar$, where the filled histograms are for the complete samples and the curve is for AMIGA without shared. The panel on the right similarly shows the lopsidedness distributions. $\HI$ line asymmetry is nearly independent of stellar mass for any given isolated sample. The two samples exhibit similar asymmetries, albeit with a small bias towards lower values for UNAM-KIAS. AMIGA shows a minor bias towards smaller stellar masses.}
    \label{alvsmstar}
\end{figure}

We plot the asymmetries against $\mstar$ in Fig.~\ref{alvsmstar}, where each sample is shown in a distinct colour. Each curve shows the median relation. For AMIGA, we show two curves, one for the full sample and another for the subsample devoid of galaxies shared with UNAM-KIAS. This is done to examine the bias introduced by the overlap. (We could have removed these galaxies from UNAM-KIAS instead, but AMIGA has better statistics.) We refer to this as ``AMIGA without shared'' hereafter. In addition,  the figure has rectangular panels at the top and right
of the main panel that show the probability distribution functions for $\mstar$ and $A_l$, respectively. AMIGA without shared is shown in these panels using curves, and the complete samples are shown as filled histograms.

The PDFs for $\mstar$ suggest that AMIGA is slightly biased towards smaller masses. Our KS-test shows that this discrepancy is at the $2\sigma$ significance level ($p$-value $\approx 0.03$) for AMIGA without shared. We also find a small systematic trend towards lower asymmetries in UNAM-KIAS, but this is not statistically significant
at a $2\sigma$ level. This is because of two reasons. First, $\HI$ line asymmetry is virtually independent of $\mstar$, which precludes the bias in $\mstar$ from reflecting in $A_l$. Secondly, the minor differences in
asymmetries at fixed mass are within the random errors.

In addition, note that galaxies with $A_l>0.1$ comprise $< 30$ per cent of either of our samples, indicating that they generally possess dynamically stable and axisymmetric $\HI$ disks \citep{Manuwal2022}. This threshold is close to the cut suggested by \citet{Bok2019} for identifying merging systems (i.e. $A_{fr}>1.26$ or $A_l>0.115$), which is satisfied by less than 24 per cent of our galaxies. These results provide further confidence that our galaxies are genuinely located in rarer environments and that their evolution is predominantly dictated by secular processes. 

\subsection{Inclination}\label{inclination} 
As the inclination increases for a disk-dominated galaxy, the $\HI$ line profile captures less of the rotational dynamics and more of the thermal and other out-of-the-plane motions (such as galactic fountains)\footnote{For an animated illustration, see \url{https://youtu.be/J0IqF1BnEJI}.}. In addition, the profile is resolved with progressively fewer velocity channels with decreasing inclination. This precludes drawing accurate physical inferences about galactic $\HI$ for face-on orientations. 
Indeed, unresolved $\HI$ emission lines are usually more asymmetric at edge-on inclinations, provided very low inclinations (that can show spuriously high asymmetries) are avoided \citep{Manuwal2022}. This does not mean that differences between galaxy populations would be entirely suppressed if inclination were not taken into account. For instance, the asymmetries of satellites exhibit a clear systematic bias towards higher values than centrals even when viewed through random lines of sight \citep[see][]{Watts2020,Watts2020b,Manuwal2022}. Minor (yet statistically significant) discrepancies, however, may not be evident to this degree unless most galaxies are appreciably inclined.

\begin{figure}
    \centering
    \includegraphics[width=1\linewidth]{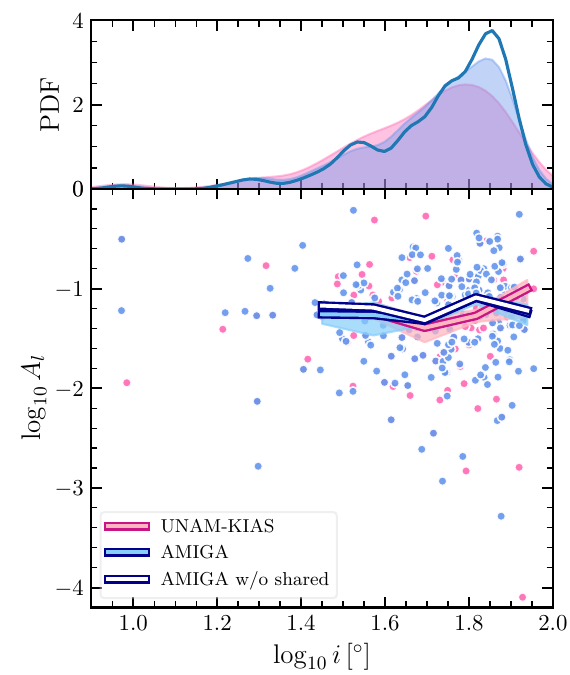}
    \caption{$\HI$ lopsidedness plotted against galaxy inclination. The asymmetry is generally independent of inclination for our isolated samples, except for a slight positive correlation at $i\gtrsim 40^\circ$ for UNAM-KIAS. The asymmetries are broadly consistent between the samples at low inclinations, but UNAM-KIAS shows slightly lower asymmetries at $60^\circ\lesssim i\lesssim 70^\circ$. UNAM-KIAS also shows a minor preference towards lower inclinations compared to AMIGA.}
    \label{alvsinc}
\end{figure}

We explore the inclination effects on the asymmetries in our $\HI$ samples in Fig.~\ref{alvsinc}. The median relations for AMIGA show that the asymmetries are agnostic to inclination, that is, there is no inclination bias. This means that the scatter in asymmetry at fixed $\mstar$ is not modulated
by inclination and rather has a physical origin, as is also hinted by the fact that asymmetry varies at fixed inclination. For UNAM-KIAS, we find a weak positive correlation ($\rho\approx0.21$)
with inclination beyond $i\approx 40^\circ$, where most of the data lies, but the $p$-value $\approx 0.05$ indicates low significance. Therefore, inclination contributes only minimally to the scatter of the asymmetry in UNAM-KIAS. This is not limited to any particular mass range, as we have confirmed that inclination is independent of mass and is also consistent between the two samples at a given $\mstar$ (not shown).

Interestingly, the asymmetries in UNAM-KIAS are consistent with AMIGA at lower inclinations, but are, on average, lower for galaxies at $60^\circ\lesssim i\lesssim 70^\circ$. This offset is apparent only at high inclinations because the $\HI$ lines capture more of the galactic velocity structure, and disappears at very high inclinations, possibly due to low statistics in UNAM-KIAS.

We would also like to point out the peculiar systems at $i\lesssim 20^\circ$ with $A_l\gtrsim 0.1$, which exhibit curiously high asymmetries for such orientations. This is not due to poor velocity resolution of their profiles, because our selection includes a cut on the number of velocity channels (Section~\ref{chanandmag}). It is reasonable to infer that the asymmetries of these objects are primarily driven by vertical motions toward or away from the disk, but the underlying mechanisms remain obscure. This makes them promising candidates for dedicated case studies with resolved $\HI$ observations in the future.

Furthermore, we note that UNAM-KIAS shows a slight bias towards lower inclinations than AMIGA, and this bias is more pronounced in AMIGA without shared. This, however, corresponds to $\Delta i\approx 5^\circ$ for the medians, and both KS and Mood's median tests show $<2\sigma$ significance levels. Hence, the bias would not be the primary driver of any strong global systematic that may exist between these samples.

\subsection{Optical morphology}\label{morph}
$\HI$ is a key governing factor in a galaxy's morphology, as it provides the rotational support required for sustaining or bringing forth a flat/disky structure, and is also intimately linked to star formation. Moreover, mechanisms that perturb $\HI$ may, if strong enough, modify the stellar distribution and global morphology. Hence, it is reasonable to expect a connection between $\HI$ line asymmetry and stellar morphology, and has even been found for centrals in \textsc{\large eagle} \citep{Manuwal2022}. In this subsection, we examine this relationship for our samples.

\begin{figure}
    \centering
    \includegraphics[width=1\linewidth]{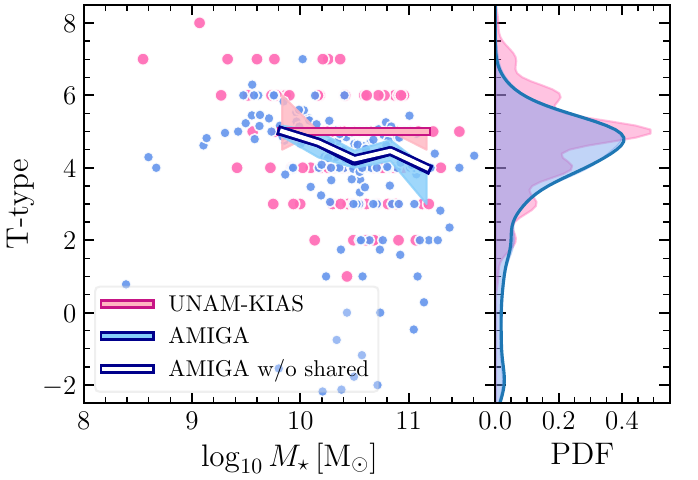}
    \caption{Optical morphology plotted against stellar mass. The morphology is independent of mass for UNAM-KIAS, but shows a negative correlation for AMIGA. Most of the galaxies are late-type spirals
    for both samples, akin to other isolated samples in the literature.}
    \label{morphvsmstar}
\end{figure}

Fig.~\ref{morphvsmstar} presents the optical morphology against $\mstar$, where the morphology has been parameterised to follow the T-type scheme in \citet{Nair2010}. The morphology distributions on the right clearly show that nearly all the galaxies span $1\leq$~T-type~$\leq8$ (i.e. Sa to Sdm), with a median of $\approx 4.5$ for AMIGA and 5 for UNAM-KIAS. Our KS and Mood's tests show that this offset between the samples, albeit small, is significant at the $>3\sigma$ level,
and has even higher significance if we exclude the AMIGA objects at T-type~$<1$. Nonetheless, the average values in both are higher than the general expectations at these masses \citep{Vazquez-Mata2022,Vazquez-Mata2025}, thereby demonstrating that our galaxies are mostly spirals, in agreement with the consensus in the literature on isolated objects \citep{Sulentic2006,Hernandez-Toledo2008,Lorenzo2012,Buta2019}. This marked preference for late-type morphology is rather expected from theory \citep{Hirschmann2013}, and is thought to arise because galaxies in rarer environments face lower likelihoods of being involved in events that reduce rotational support and transform them into more elliptical structures, like gas stripping, mergers, and merger-induced AGN feedback \citep[see][]{Pfeffer2023}. 

The figure also shows that T-type is negatively
correlated for AMIGA (Spearman rank test $\rho\approx -0.36$ and $p$-value~$\sim 10^{-6}$). This behaviour may appear qualitatively similar to the typical mass-dependence of morphology, but exhibits a key difference in that massive isolated galaxies ($\mstar\gtrsim 10^{10.7}\,\msol$) are mostly spirals, not ellipticals. We find a similar trend for UNAM-KIAS ($\rho\approx -0.19$), albeit at a $2\sigma$ significance level. This is not apparent in the median relation. Adding random errors of $\lesssim 0.5$ to UNAM-KIAS points yields a negative slope below $\mstar\sim 10^{10.5}\,\msol$, but does not change the flattening at higher masses. Therefore, the flattening below this mass is likely an artifact caused by the highly discretised nature of the data, but is due to the larger fraction of galaxies with T-type~$\gtrsim 5$ at higher masses.

An important caveat of the results presented above is that morphological classifications are more robust at lower inclinations. This is because it is easier to discern galactic features that would otherwise be missed, such as bars, rings, and lenses \citep[e.g.][]{Nair2010}, and due to the artificial reweighting of different components caused by dust attenuation and projection \citep{Driver2007,Pastrav2013}. Inclinations below $\approx 50-60^\circ$ are generally considered safe for reliable assessments \citep{Marinova2007,Barazza2008,Aguerri2009,Nair2010,Vazquez-Mata2022,Vazquez-Mata2025}. We impose upper limits of $i\lesssim 40^\circ-60^\circ$ and find that the aforementioned results, despite their lower significance levels, generally hold true. The only major change is for the negative trend between morphology and mass in UNAM-KIAS, which disappears after enforcing the thresholds, but this could also stem from poor statistics as we are left with just 72 objects.

\begin{figure}
    \centering
    \includegraphics[width=1\linewidth]{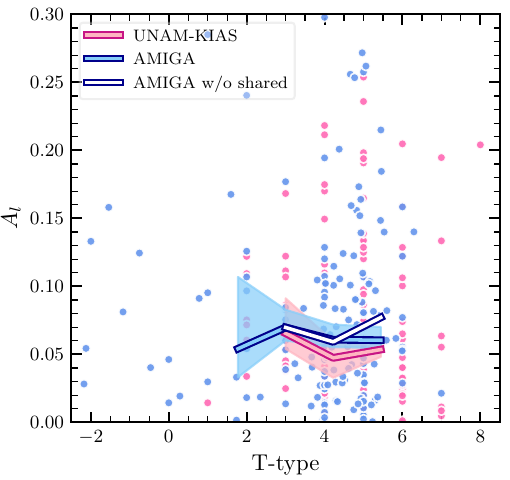}
    \caption{$\HI$ line asymmetry plotted against stellar morphology (T-type). The asymmetry is agnostic
    to morphology and consistent between the samples for a given T-type.}
    \label{alvsmorph}
\end{figure}

Next, we examine the relationship between $\HI$ line asymmetry and optical morphology in Fig~\ref{alvsmorph}. The median relations show that the asymmetry is independent of stellar morphology for both UNAM-KIAS and AMIGA, and that it is the same for a fixed morphology across samples, at least for the bins with sufficient statistics. We have also confirmed this through Spearman rank correlation tests. Additionally, we conduct a KS test to compare the asymmetries of galaxies lying above and below T-type~$=1$ in AMIGA, marking the transition from early- to late-types, and find that the two morphological groups are virtually identical in this respect.

Both of these quantities are generally sensitive to inclination, but in opposite ways: $A_l$ becomes more reliable at high inclinations, whereas T-type is less reliable. This means there is an intermediate $i$ range where both carry similar degrees of reliability. We assume $30^\circ<i<70^\circ$ as a suitable range for our purpose and re-examine the correlations, but do not find any trends. The results remain the same even after modifying the limits by $\pm 10^\circ$.

\subsection{Stellar asymmetry}
We now extend our analysis to explore stellar asymmetries within our samples by using two kinds of spatial optical asymmetry measures: rotational asymmetry ($A$) and shape asymmetry
($A_s$). The latter is similar to $A$ but is computed using a binary mask that identifies pixels
dominated by galactic emission, making it more sensitive to low surface brightness features on the outskirts than $A$ \citep[see][]{Pawlik2016,Rodriguez-Gomez2019}. On the other hand, since $A_s$ is agnostic to light contrasts associated with spiral arms, $A$ may sometimes be better at capturing asymmetries associated with such structures. 

\begin{figure}
    \centering
    \includegraphics[width=1\linewidth]{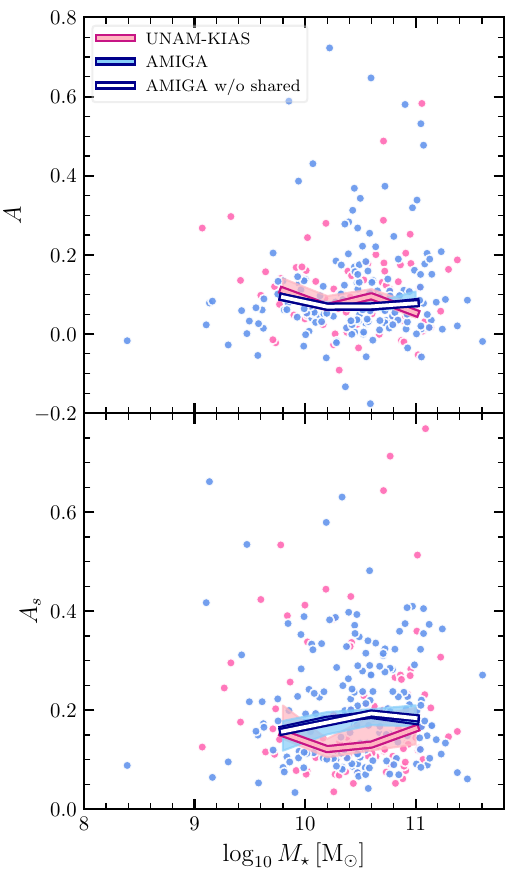}
    \caption{Spatial asymmetry of stars plotted against stellar mass. The top panel shows the results for the rotational asymmetry ($A$) and the bottom panel for the shape
    asymmetry ($A_s$), the latter being more sensitive to the outskirts of the galaxy (see the text). The stellar asymmetry is generally independent of stellar mass for both measures, but there
    is a mild positive relation in UNAM-KIAS for $\mstar\gtrsim 10^{10.2}\,\msol$. At fixed mass, UNAM-KIAS shows slightly lower $A_s$ than AMIGA.}
    \label{asvsmstar}
\end{figure}

These asymmetries are shown for our samples in Fig.~\ref{asvsmstar} for different stellar masses, where the top and bottom panels show the results for $A$ and $A_s$, respectively. One may notice that $A$ is negative
for some galaxies. This is because the asymmetry is dominated by the background term in the definition of $A$. Nevertheless, these cases are only $\approx 10$ per cent of either sample, and we have confirmed that the trends discussed below are robust to whether these peculiar measurements are included or not.

A comparison of the median values in the top and bottom panels shows that $A_s\gtrsim A$
in general, as expected from their definitions. The top panel, in particular, shows that $A$ is flat with respect to mass and is identical across the samples on average. This is similar
to the behaviour for $A_l$ seen earlier. We find no overall trend for $A_s$ either, as is apparent
in the bottom panel. We note, however, that UNAM-KIAS does not exhibit a flat relation but rather a U-shaped median curve, indicating that there are two opposite trends in the data that switch at $\mstar\sim 10^{10.2}\,\msol$. Indeed, our Spearman rank test reveals a
mild positive correlation above this mass with $\rho\approx 0.21$ ($p$-value~$\approx 0.08$).
Because these measurements are not accurate near edge-on inclinations, we also examined these results
for $i<60^\circ$. We find that the trends hold and that the positive correlation at higher masses seen
for $A_s$ in UNAM-KIAS is in fact stronger ($\rho\approx 0.29$). 

Additionally, we find that UNAM-KIAS typically shows a lower median $A_s$ than AMIGA at the same mass. This is not an inclination effect, as UNAM-KIAS is biased towards slightly lower inclinations (Fig.~\ref{alvsinc}), which should enable better assessments of spatial asymmetry than AMIGA and yield higher values instead. The fact that $A$ does not show any offset indicates that the two mainly differ in asymmetry levels at larger radii.

\begin{figure}
    \centering
    \includegraphics[width=1\linewidth]{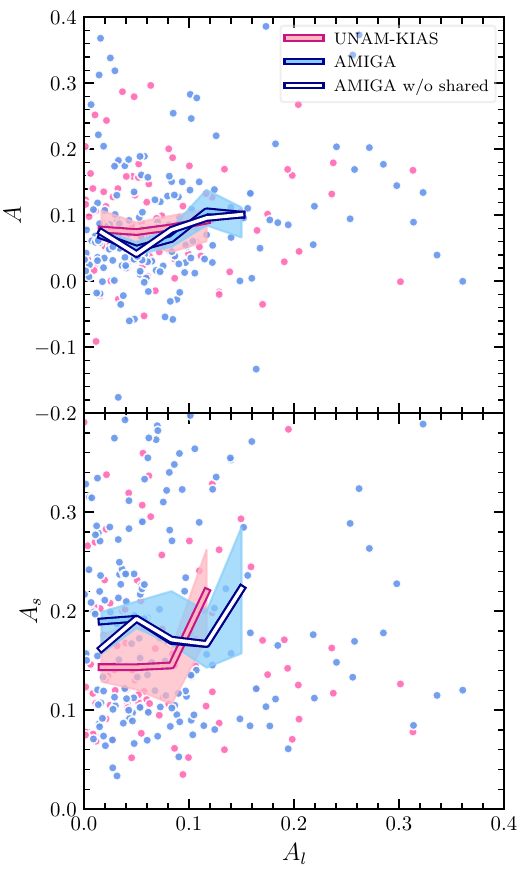}
    \caption{The relationship between spatial asymmetry of stars and $\HI$ line asymmetry. The top panel shows the results for the rotational asymmetry ($A$), and the bottom panel shows it with the shape asymmetry ($A_s$). $\HI$ line asymmetry is independent of optical asymmetry for UNAM-KIAS, but $A$ exhibits a weak positive correlation for AMIGA.}
    \label{alvsas}
\end{figure}

We now turn to Fig.~\ref{alvsas} that shows the relationship between $A_l$ and the two
stellar asymmetries. The top and bottom panels again carry the same meaning as in the previous figure. The top panel shows that $A$ is nearly constant in $A_l$ for UNAM-KIAS, but there is a minor positive trend in AMIGA (Spearman rank $\rho\approx 0.17$ and $p$-value~$\approx 0.02$). We have confirmed that this trend persists even after imposing reasonable inclination limits, where both stellar and $\HI$ measures are sufficiently reliable ($30^\circ<i<60^\circ$, for example). On the other hand, no correlation for $A_s$ is observed in either sample, whether we consider the whole sample, some specific mass range, or impose inclination limits.

Although the trend in $A$ for AMIGA is weak, it is significant enough to ask a pertinent question: why is there no trend in $A_s$? Given that $A$ is, by construction, light-weighted and biased towards inner regions, the trend at higher masses in $A_s$ and its absence in $A$ can be interpreted as indicating that $\HI$ asymmetry in AMIGA is correlated with stellar asymmetry in the inner regions of the galaxy but not with that in the outer regions. This is contrary to expectations because $\HI$ is far more extended than the stellar disk. However, although $A_s$ is better at capturing asymmetries at high radii, its insensitivity to variations in surface brightness can sometimes prevent it from gauging asymmetries in galactic structures like spiral arms. This information is better captured by $A$, which could explain the absence of trends in $A_s$. A proper confirmation would require further analysis that is beyond the scope of this work.

\subsection{$\HIb$ fraction}\label{hifractions}
$\HI$ content is a useful probe of gas flows to and from a galaxy. It is also expected to exhibit some correlation with asymmetry because gas flows modify the $\HI$ phase space. For instance, because gas accretion can impart turbulence to cold gas disks \citep{Bournaud2005,Heald2011,Chung2012,Forbes2023,Jimenez2023}, we expect to see a positive correlation between $M_{\HIs}$ and asymmetry if it is the main driver. Alternatively, a negative correlation between these properties is expected if the disturbance is primarily induced by a process that depletes $\HI$, like stellar/AGN feedback or tidal stripping. Since $M_{\HIs}$ correlates with $\mstar$, a more suitable and prevalent parameter for exploring these effects is $M_{\HIs}/\mstar$, i.e. the $\HI$ fraction \citep[e.g.][]{Watts2020,Reynolds2020,Watts2021,Glowacki2022}. In this section, we examine the $\HI$ fractions of our isolated galaxies and their relationship to the $\HI$ line asymmetry.

Prior works have shown that AMIGA and UNAM-KIAS exhibit a clear deviation towards higher $\HI$ fractions relative to the general population \citep{Bradford2015,Calette+2018,Bok2019}. One reason for this is that isolated galaxies are almost exclusively centrals, and 
centrals are relatively immune to environmental gas removal and more efficient at gas accretion than satellites \citep{Cen2014,Sawala2012,Voort2017,Stevens2019}. In fact, isolated galaxies are often used to quantify the $\HI$-deficiencies of galaxies in denser environments \citep{Haynes1984,Solanes2001,Verdes-Montenegro2007,Boselli2014,Denes2014,Grossi2016}. However, as we soon demonstrate, this does not suffice to explain the full extent of the offset.

\begin{figure}
    \centering
    \includegraphics[width=1\linewidth]{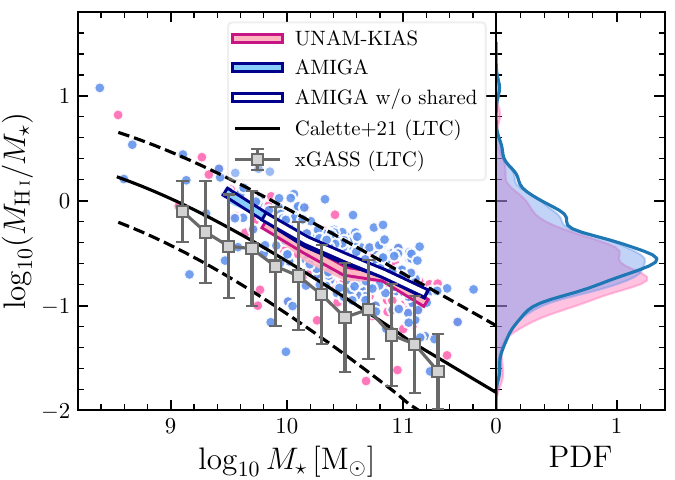}
    \caption{$\HI$ fractions plotted against stellar mass for our isolated samples. Overplotted is the mean relation and the scatter for late-type centrals (LTC) derived by \citet{Calette+2021a} as the solid black curve, along with the mass-dependent standard deviations shown using the dashed curves. The grey points and their error bars show the mean $\HI$ fractions and $1\sigma$ scatter for the LTC in xGASS after including the upper limits for non-detections. Our galaxies are generally richer in $\HI$ than typical late-type objects. UNAM-KIAS exhibits slightly lower $\HI$ fractions than AMIGA.}
    \label{hifracvsmstar}
\end{figure}

We show the $\HI$ fractions of our galaxies in Fig.~\ref{hifracvsmstar}. In addition, we plot the empirical mean relation (solid curve) and the standard deviations at different masses (dashed black curves) for late-type centrals (hereafter LTC) provided by \citet{Calette+2021a}. This was obtained from a conditional analytic distribution for late-type galaxies derived by \citet{Calette+2018} via a thorough analysis of a homogenised compilation of multiple, high-quality datasets from the literature (including xGASS), accounting for non-detections through a survival analysis procedure. The distribution for centrals was constrained using the central/satellite classifications in the xGASS catalog that were in turn borrowed from the \citet{Yang2007} group catalog for SDSS. For reference, we also plot the values for the LTC in xGASS (squares with error bars) obtained with the upper limits for non-detections. xGASS is fairly consistent with the \citet{Calette+2021a} distribution, which is expected given that the subsample of late-type galaxies is dominated by detections.

Interestingly, our isolated galaxies are not only $\HI$-rich but also exhibit higher $\HI$ fractions than typical LTC, with most of them nonetheless within the observed scatter. This means that they simply represent the most $\HI$-rich objects within the broader population of LTC. The offset in the median $\HI$ fraction relative to the solid black curve decreases with mass for $\mstar\lesssim 10^{10.2}\,\msol$ and increases thereafter, reaching $\approx 0.5$~dex at $\mstar\sim 10^{11.2}\,\msol$. Similar results are found with xGASS. Although not shown here, we also find a preference for higher $\HI$ fractions when comparing with ALFALFA \citep[see][]{Watts2021}, but with smaller offsets than with xGASS. This is rather expected given that ALFALFA is well known to be highly biased towards $\HI$-rich galaxies.

Note that both of our samples are $\HI$-rich compared to typical LTC, but they are not exactly equivalent in detail. The $\HI$ fractions in UNAM-KIAS are systematically lower than in AMIGA by $\lesssim 0.1$~dex, a discrepancy that our KS and Mood's tests show to be significant at the $\gtrsim 2\sigma$ level and stronger with AMIGA without shared. We recommend that the reader keep this bias in mind when using our samples as references to study the gas reservoir of galaxies in other environments.

For the $\HI$ mass to differ between two galaxies, they must either accrete or deplete at different rates, or both. This means that investigating the contribution from these phenomena can shed light on the physics underpinning the results in Fig.~\ref{hifracvsmstar}. We attempt this in the subsequent portion of the subsection, where we start by addressing the higher-than-usual $\HI$ masses of our galaxies, and then discuss the differences between our UNAM-KIAS and AMIGA samples. To this end, we first investigate the contribution from gas accretion onto our galaxies. Although we do not possess direct measurements of accretion rates, insights from theoretical works can aid in indirect assessments.

The conventional picture holds that gas falling into a massive halo undergoes shock heating to temperatures $T\gtrsim 10^{5}\K$ and develops a quasi-static hot halo supported by thermal pressure \citep{Rees1977,Silk1977,White1978,White1991,Birnboim2003}. Cold gas is fed to the central galaxy via radiative cooling in the inner regions of the hot halo and/or cosmological cold streams, with the relative dominance of these modes varying with halo mass and redshift \citep{Keres2005,Dekel2006,Brooks2009,Voort2011,Nelson2013,Correa2018a,Wright2021,Waterval2025}. However, feedback from stellar or AGN activity can complicate this picture. For instance, not all gas brought in via these modes may reach the galaxy due to the disruption of gas inflows \citep{Nelson2015,Correa2018b}. On the other hand, feedback ejecta can also induce cooling of the hot corona after phase-mixing \citep[e.g.][]{Marinacci2010,Barbani2023}. Despite these complexities, simulations have shown that the gas accretion rate is mainly determined by the halo mass, involving a steeper rise up to $M_{\rm halo}\approx 10^{12}\,\msol$ and a milder rise at higher masses \citep{Voort2011,Correa2018b,Mitchell2020b}.

\begin{figure}
    \centering
    \includegraphics[width=1\linewidth]{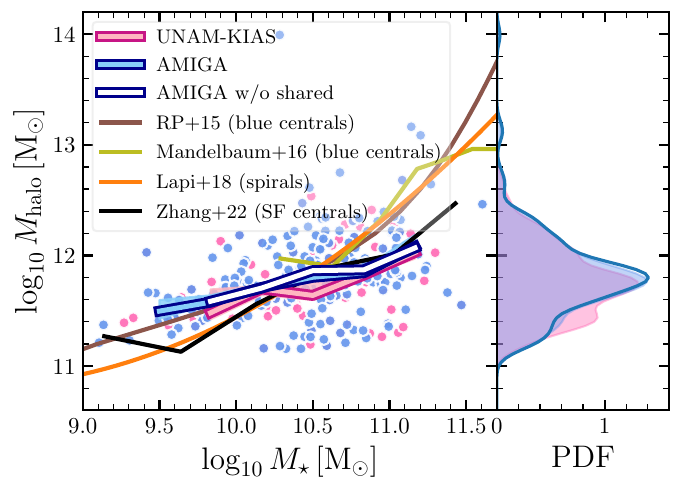}
    \caption{Stellar-to-halo mass relations (SHMRs) for our samples. The brown curve is the relation for blue centrals by \citet{Rodriguez-Puebla2015} based on abundance matching. The olive curve is the relation for blue centrals derived by \citet{Mandelbaum2016} using weak lensing. The orange curve is the relation inferred from rotation curves of local spirals by \citet{Lapi2018}. The black curve is the relation from \citet{Zhang2022} for SF centrals using weak lensing. The isolated galaxies at $\mstar\lesssim 10^{10.3}\,\msol$ reside in haloes that are typical of blue centrals, but are more massive than those hosting SF centrals or spirals. At higher stellar masses, the halo masses are similar to SF centrals but lower than blue/spiral galaxies. UNAM-KIAS and AMIGA are consistent in their SHMRs in this plane within the statistical uncertainties.} 
    \label{shmr}
\end{figure}

We show the stellar-to-halo mass relations (SHMRs) for our samples in Fig.~\ref{shmr}, where the halo masses are taken from the \citet{Tinker2021} group catalog, one of the few catalogs that account for the SFR-dependence of the SHMR. In addition, we overplot the SHMRs for blue centrals by \citet{Rodriguez-Puebla2015} and \citet{Mandelbaum2016}, the relation for local spirals by \citet{Lapi2018}, and that for SF centrals by \citet{Zhang2022}. These relations rely on different methods for halo mass measurements: \citet{Mandelbaum2016} and \citet{Zhang2022} utilise weak lensing, \citet{Lapi2018} use rotation curves, and \citet{Rodriguez-Puebla2015} populate simulated haloes with galaxies using a semi-empirical (abundance matching) approach. 

Before we compare our samples with any of these relations, it is important to note the discrepancies among them at $\mstar\lesssim 10^{10.3}\,\msol$ and $\mstar\gtrsim 10^{10.7}\,\msol$. One of the main reasons for this is sampling bias. Although galaxy colour, morphology, and SFR are well-correlated quantities, they do not exactly capture the same physical information and also evolve at different timescales \citep[e.g.][]{Correa2019,Davies2020}. In fact, the apparent discrepancy between \citet{Zhang2022} and \citet{Mandelbaum2016} vanishes if the former SHMR is produced using the latter's colour selection \citep[see][]{Zhang2022}. This also explains the good agreement between \citet{Rodriguez-Puebla2015} and \citet{Mandelbaum2016}. Additional differences may arise because of the varied approaches used in these works for stellar or halo mass estimation. This calls for caution when using these relations interchangeably.

Notwithstanding, we find that our samples are always consistent with at least one of these relations at any given $\mstar$.
At face value, the results in Fig.~\ref{shmr} suggest that the isolated galaxies are more representative of blue centrals at $\mstar\lesssim 10^{10.3}\,\msol$ and of SF centrals at $\mstar\gtrsim 10^{10.7}\,\msol$. We believe, however, that it is more reasonable to use \citet{Lapi2018} or \citet{Zhang2022} as the reference given the morphologies (Fig.~\ref{morphvsmstar}) and SFRs of our galaxies (Fig.~\ref{ssfrvsmstar}). Doing so reveals that, at $\mstar\lesssim 10^{10.3}\,\msol$, our galaxies are hosted by haloes with masses $\lesssim 0.4$~dex higher. For the halo masses covered by the SHMRs at these $\mstar$s, simulations indicate that the (absolute) gas accretion rate onto the central galaxy increases, on average, by $\approx 1.3-1.5$~dex for every increase in $M_{\rm halo}$ by a dex \citep{Correa2018b,Mitchell2020b}. This implies that our galaxies are accreting cold gas at rates that are higher than those for typical SF/late-type galaxies by $\lesssim 0.6$~dex. Therefore, higher gas accretion rates may contribute to the unusually high $\HI$ masses of our galaxies below $\mstar\sim 10^{10.3}\,\msol$ (Fig.~\ref{hifracvsmstar}). This does not apply to higher stellar masses because
the halo masses are consistent with \citet{Zhang2022} and lower than those in \citet{Lapi2018}. Note, however, that gas loss can, in principle, compensate for the impact of gas accretion at lower masses. This is what we explore next.  

\begin{figure}
    \centering
    \includegraphics[width=1\linewidth]{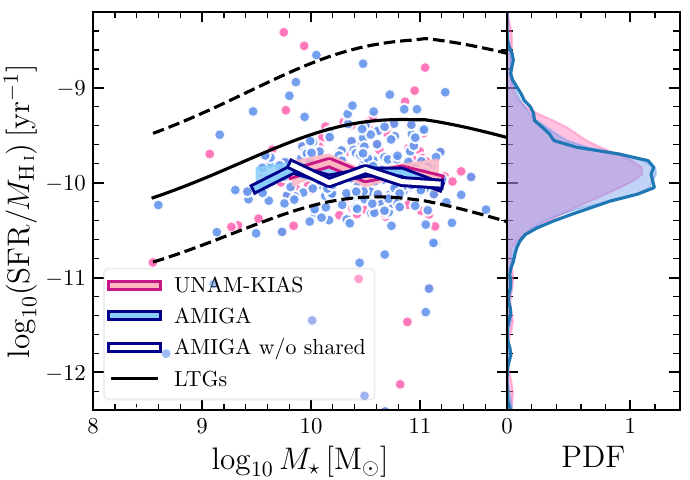}
    \caption{Star formation efficiency (SFE) based on $\HI$ mass at different stellar masses. For reference, we have added the means and $1\sigma$ scatters for LTC calculated using the $\HI$ fractions and the SFMS in \citet{Calette+2021a} and \citet{Stephenson+2024},
    respectively. The SFE is agnostic to stellar mass. UNAM-KIAS and AMIGA are biased towards lower SFEs than typical SF galaxies. The two are consistent in their median SFEs, but UNAM-KIAS exhibits a larger scatter.}
    \label{sfevsmstar}
\end{figure}

For the mass range explored here, a central galaxy can lose $\HI$ mainly via two channels: conversion to $\Hmol$ and the subsequent consumption via star formation, and feedback-driven ejection. We assess the role of gas consumption through the star formation efficiency based on $\HI$ content, ${\rm SFE}={\rm SFR}/M_{\HIs}$. These are shown for our samples in Fig.~\ref{sfevsmstar}. We also estimate the SFEs expected for typical LTC using the $\HI$ fractions by \citet{Calette+2021a} and the star-forming main sequence (SFMS) for the central galaxies by \citet{Stephenson+2024} (shown in Fig.~\ref{ssfrvsmstar}). The resultant mean SFE is shown as the solid black curve and the (mass-dependent) $1\sigma$ scatter is shown using the dashed curves. The figure shows that our samples are biased towards SFEs $\approx 0.4-0.6$~dex lower than most LTC, implying slower depletion of their gas reservoirs. 

The results in Figs.~\ref{hifracvsmstar} and~\ref{sfevsmstar} altogether show that the isolated galaxies are $\HI$-rich but do not consume $\HI$ intensively, which is expected when the consumption is regulated only by stellar feedback: supernovae heat the ISM and increase the ISM velocity dispersion, keeping the Toomre instability parameter \citep{Toomre1964} around the critical regime ($Q_g\sim1-2$; \citealt{Firmani-Avila2000,Hopkins2011,Hopkins2012,Becerra2014}). %\adi{\citet{Firmani-Avila2000} is a valid citation here, but note that it does not strictly predict a stable $Q$ and \textit{effectively} enforces this. This is because it directly links star formation to $Q\leq 2$, and models energy dissipation in order to counter the associated increase in velocity dispersion (and hence $Q$) and bring $Q$ back to the threshold value. Regardless, the actual predictions of the model are observables, and they do present good agreement with measurements, which indirectly validates the assumption of a stable $Q\sim 2$.}
Isolated galaxies can maintain long-term stability in this regime because they are relatively immune to the torques induced during galaxy--galaxy interactions, primarily exerted by stars within the galaxy on the gas due to stellar asymmetries generated by tidal forces \citep{Hopkins2009}; this behaviour is robust against large variations in the parameterisations of star formation and feedback in simulated galaxies \citep[e.g.][]{Hopkins2011,Hopkins2012}. In denser environments, the aforementioned torques cause the $\HI$ to lose its angular momentum and flow towards the nuclear region, increasing the gas density and SFR there \citep{Matteo2007,Hopkins2009,Hopkins2013,Moreno2015,Blumenthal2018}. Indeed, observations show that paired and post-merger galaxies exhibit enhanced SFRs \citep{Rownd1999,Wong2011,Ellison2013,Patton2013,Knapen2015,Thorp2019,Pearson2025} and diluted metallicities in their central regions \citep{Kewley2006,Ellison2008,Scudder2012,Ellison2018,Thorp2019}. Furthermore, even if isolated galaxies were to experience occasional interactions, the internal torques are predicted to be inefficient due to their high gas fractions, as there are not many stars to do the torquing \citep[see][]{Hopkins2009}. There can still be some SFR enhancement at the outer radii where the gas shocks between the two interacting objects, due to the associated density contrasts and gas compression, but this is expected to be a relatively minor effect \citep{Hopkins2013}.

Next, we focus on outflows from our galaxies. Although we do not have outflow rate measurements for our galaxies, we can infer them indirectly. We know that ISM's metallicity is governed by the interplay between enrichment associated with star formation, dilution due to gas accretion, and ejection via outflows. Moreover, SFR governs both the stellar mass growth and the rate of energy and/or momentum injection from feedback, and it is itself related to the inflow rate. It therefore follows that one can formulate a generic analytical model that quantifies these processes under reasonable assumptions, and then use the metallicity--SFR--$\mstar$ relation [i.e. the fundamental metallicity relation or FMR (\citealt{Lara-Lopez2010,Mannucci2010}); but see \citealt{Sanchez+2019})] to constrain the free parameters, and by extension, the precise scaling between sSFR, $\mstar$, and outflow rate. 

This was recently attempted by \citet{Lin2023}, who developed a flexible, non-equilibrium chemical evolution model that can be easily constrained for any galaxy sample using the corresponding FMR. By testing it on SF galaxies in \textsc{\large eagle}, the authors showed that the model simultaneously reproduces the evolution of gas-phase metallicity, SFR, and the mass loading factor ($\eta$, i.e. the outflow rate to SFR ratio). When constrained using the observed FMR in SDSS, it accurately predicts the latest down-the-barrel $\eta$ measurements within reasonable uncertainties. The key feature of this model for our purposes is that it parameterizes $\log \eta$ as a product of power laws of $\mstar$ and sSFR. The best-fit model for SDSS shows a positive exponent for sSFR, indicating that the outflow rate increases with sSFR at a fixed $\mstar$. Within this context, Fig.~\ref{ssfrvsmstar} suggests that our galaxies have been experiencing weaker outflows than most galaxies on the SFMS. 

In summary, Figs.~\ref{shmr} and~\ref{sfevsmstar} collectively indicate that our galaxies have higher $\HI$ masses due to lower SFE, weaker outflows, and higher gas accretion rates, with the latter relevant only for $\mstar\lesssim 10^{10.3}\,\msol$. Similarly, we can attempt to uncover the physical reasons for the systematically lower $\HI$ masses in the UNAM-KIAS sample compared to the AMIGA sample. 

We find no systematic difference in sSFR or SFE among these samples, suggesting that the $M_{\HIs}$ offset is mainly due to accretion rates. Fig.~\ref{shmr} appears to show slightly lower halo masses for UNAM-KIAS at $\mstar\sim 10^{10.5}\,\msol$, which also corresponds to the maximum offset in $M_{\HIs}$, but this is within the statistical errors. Moreover, unlike $M_{\HIs}$, there is no global $M_{\rm halo}$ discrepancy between the samples. This may seem puzzling because the accretion rate is mainly governed by halo mass, but note that it also has a secondary dependence on the local density at fixed halo mass \citep{Voort2017}. Taking this into account, we can infer that UNAM-KIAS galaxies have been accreting gas at slightly lower rates because they are located in denser regions than AMIGA. This conclusion is corroborated by the stricter isolation strategy adopted for the latter (as discussed in Section~\ref{amiga}).

\begin{figure}
    \centering
    \includegraphics[width=1\linewidth]{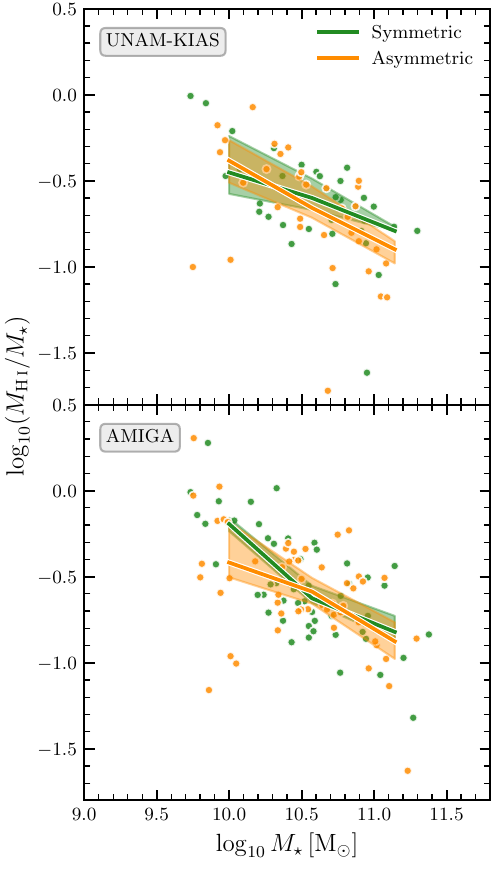}
    \caption{The relationship between $\HI$ lopsidedness and $\HI$ fraction for UNAM-KIAS (top) and AMIGA (bottom). The orange and green curves show the median values for the most and least asymmetric galaxies in each mass bin, respectively (see the text). $\HI$ fraction is generally independent of $A_l$ for both samples.}
    \label{hifracandal}
\end{figure}

We next investigate whether the $\HI$ fraction correlates with $A_l$. For this, we divide the subsample within each mass bin into the least and most asymmetric, based on whether they lie below the 25th percentile or above the 75th percentile of $A_l$, and compare the median $\HI$ fractions. The results are shown in Fig.~\ref{hifracandal}, where the top and bottom panels correspond to UNAM-KIAS and AMIGA, respectively. The median $\HI$ fractions for the symmetric and asymmetric subsamples are shown with the green and orange curves, respectively. We show these only for bins with at least 32 points, ensuring that the medians are based on $\geq 8$ points, an ad hoc threshold that provides a reasonable compromise between statistics and reliability for our data sets.

It is clear that the $\HI$ fraction is independent of $A_l$
for both of our samples, at least for the mass bins with sufficient statistics. This can occur in one of two scenarios: either the gas accretion and consumption rates are uncorrelated with asymmetry, or they both correlate to nullify any net trend between the $\HI$ fraction and asymmetry. We examine these scenarios by performing exercises similar to the one above for SFE and $M_{\rm halo}$, which reveal that SFE is agnostic to the asymmetry
and so is $M_{\rm halo}$ for AMIGA (not shown). The latter may also apply to UNAM-KIAS, but we do not have enough statistics to be conclusive.

\subsection{Star formation activity}\label{ssfr}
\begin{figure}
    \centering
    \includegraphics[width=1\linewidth]{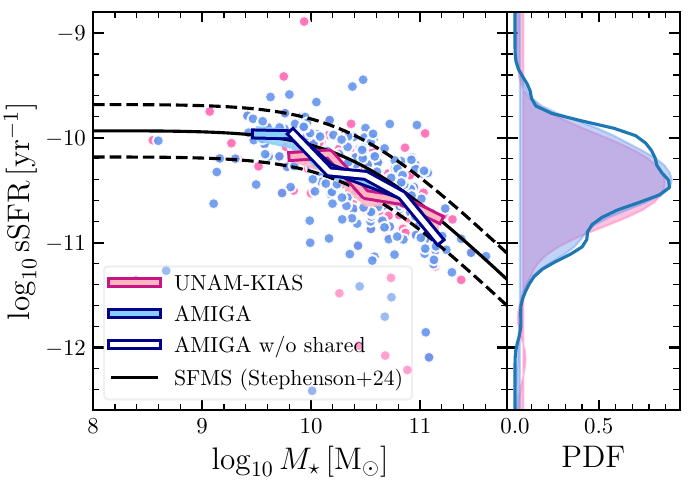}
    \caption{Specific star formation rate vs stellar mass. The black curve shows the mean star-forming main sequence (SFMS) determined by \citet{Stephenson+2024} for the central galaxies in SDSS. The dashed curves encompass $\pm0.25$~dex around the SFMS and denote the main sequence region. Our isolated galaxies are generally on the main sequence but exhibit a bias towards sSFRs below the mean SFMS. UNAM-KIAS and AMIGA are consistent in their sSFRs at fixed $\mstar$.}
    \label{ssfrvsmstar}
\end{figure}

We now examine the star-formation activity of our galaxies, quantified as the sSFR. The median sSFRs at different masses are shown in Fig.~\ref{ssfrvsmstar}. For reference, we also include the mean star-formation main sequence (SFMS) by \citet{Stephenson+2024}, shown as the solid black curve. To obtain this SFMS, the authors used GSWLC SFRs and $\mstar$s, making it suitable for fair comparisons with our samples. Fig.~\ref{ssfrvsmstar} also includes two dashed curves, offset by $\pm 0.25$~dex from the mean SFMS, and enclose the total SFMS \citep[e.g.][]{Salim2007,Whitaker2012,Speagle2014}.

Our galaxies are mostly within the SFMS, but they are not distributed uniformly and show a preference for sSFRs below the mean for both UNAM-KIAS and AMIGA. These two samples are consistent in their sSFRs at a given $\mstar$, as shown in the right panel. This bias is expected given the $\HI$ fractions and SFEs presented in the previous subsection, and occurs because these disk-dominated galaxies are forming stars in the quasi-equilibrium regime characterised by self-regulated star formation (as elaborated in Section~\ref{hifractions}).

Note that the SFRs used to generate these trends probe star formation over long timescales $\sim 100\,{\rm Myr}$\,\,\citep{Calzetti2013,Flores+2021}, unlike those based on tracers that capture more instantaneous activity (such as ${\rm H\,\alpha}$). The results shown here are therefore not strongly influenced by stochastic fluctuations in star formation activity. 

\begin{figure}
    \centering
    \includegraphics[width=1\linewidth]{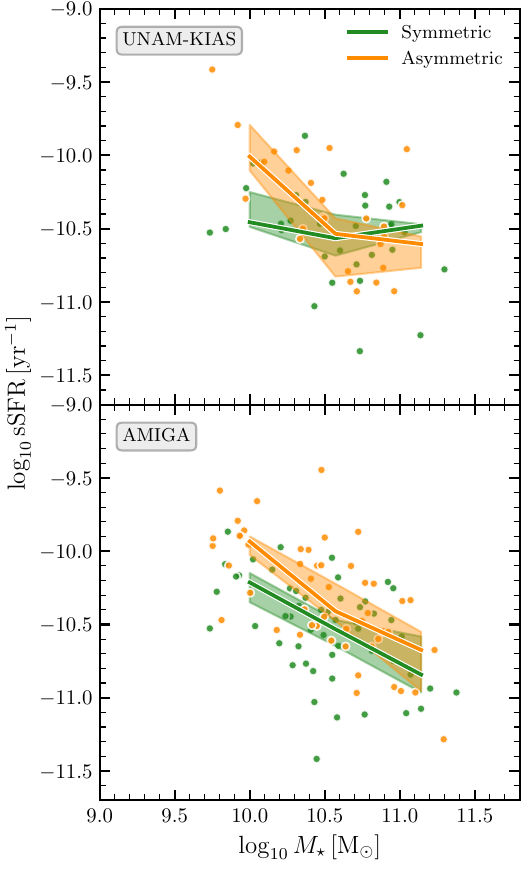}
    \caption{Specific star formation rates for the least and most $\HI$ asymmetric galaxies in UNAM-KIAS (top panel) and AMIGA (bottom). The orange curve shows the median asymmetry for upper quartile of $A_l$ within each mass bin, and the green curve shows it for the lower $A_l$ quartile. Asymmetric galaxies show higher sSFRs below $\mstar\sim 10^{10.3}\,\msol$.}
    \label{ssfrandal}
\end{figure}

We now investigate how sSFR relates to $\HI$ line asymmetry within our samples. Since sSFR depends on $\mstar$, we need to examine differences at a given mass. For this, we divide the galaxies in each mass bin into 25 per cent least and most asymmetric galaxies based on the $A_l$ distribution for that bin and compare the median asymmetries in these two subsamples. The results of this exercise are shown in Fig.~\ref{ssfrandal}, where the top and bottom panels correspond to UNAM-KIAS and AMIGA, respectively. In each panel, the orange points indicate the most asymmetric galaxies
at that mass, and the green points show the symmetric counterparts. 

The top panel shows that, for both UNAM-KIAS and AMIGA, $\HI$ fractions are agnostic to the $A_l$ if $\mstar\gtrsim 10^{10.3}\,\msol$, but asymmetric galaxies show higher sSFRs at lower masses. Interestingly, \citet{Manuwal2022} analysed this relationship for the central galaxies in the \textsc{\large eagle} simulation and also found that asymmetric objects prefer higher sSFRs
at $\mstar\lesssim 10^{10.3}\,\msol$, and that there is a switch in this relationship at higher masses. However, they found this to correspond to a reversal in the trend rather than the independence seen for the massive galaxies in our samples. The authors used differences in gas outflow rates at different masses combined with their sSFR--$A_l$ trends to infer that
the asymmetries in centrals at $\mstar\lesssim 10^{10.3}\,\msol$ are likely caused
by outflows driven by stellar feedback, whereas those at higher masses are linked to AGN feedback. Our results for isolated galaxies are partially consistent with this interpretation for $\mstar\gtrsim 10^{10.3}\,\msol$, in that $\HI$ asymmetries at these masses are not driven by stellar feedback for either UNAM-KIAS or AMIGA. Recall that sSFR is a good proxy for feedback-driven outflow rates in observed galaxies (see Section~\ref{hifractions}), which implies that the positive correlation at lower masses between sSFR and asymmetry is likely caused by varying outflow rates, just as predicted by \citet{Manuwal2022}.

\subsection{Bar presence}\label{bars}
Isolated galaxies are prone to spontaneous bar formation due to gravitational instabilities within the disk \citep[for details, see the review by][]{Sellwood2014}. Although the strength and growth of this bar depend on the initial gas content of the disk, they are also sensitive to other factors such as bulge properties, disk thickness, velocity anisotropy, halo mass and shape \citep[e.g.,][]{Klypin2009,Villa-Vargas2010,Athanassoula2013,Lokas2020}. On the other hand, bars can modify the $\HI$ phase-space and drive radial inflows that build a central gas concentration and enhance star formation \citep{Fanali2015,Verwilghen2024}, or prevent star formation by introducing turbulence in the gas \citep{Khoperskov2018,Rosas-Guevara2020}. This becomes even more complicated if the galaxy accretes gas or interacts with neighbouring galaxies \citep{Bournaud2002,Lokas2016,Cavanagh2020,Lokas2021,Fragkoudi2025}. Needless to say, bars and $\HI$ exhibit a complex causal relationship that is not yet fully understood, but one that must be unravelled to advance our understanding of bar formation and its connection to galaxy evolution. Observed samples of isolated galaxies with bar information carry considerable utility in this regard.

\begin{figure}
    \centering
    \includegraphics[width=1\linewidth]{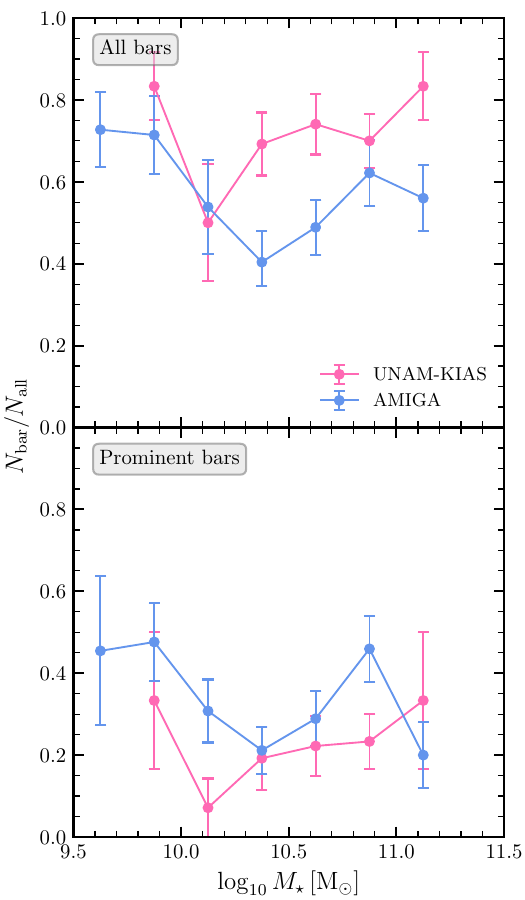}
    \caption{Fraction of galaxies with bars at different masses. The error bars are the uncertainties on the fractions from bootstrapping. The top and bottom panels show the results for all kinds of bars and strong/prominent bars, respectively. The bar fraction tends to be higher for UNAM-KIAS in general, but is slightly lower for strong bars.}
    \label{fbarvsmstar}
\end{figure}

We show the incidence of bars in our galaxies in Fig.~\ref{fbarvsmstar}, which plots
the fraction of galaxies with bars in different stellar mass bins. The error bars show the
variations predicted by bootstrap resampling. Again, we only plot these fractions for the bins with $>=8$ points. The top panel shows the results for all types of bars, while the bottom panel focuses only on prominent (or strong) bars.

The global bar fractions (top panel) show minima at $\mstar\sim 10^{10.1}\,\msol$
for UNAM-KIAS and at $\mstar\sim 10^{10.4}\,\msol$ for AMIGA. 
This trend is qualitatively similar to that observed in previous works using SDSS (or similar) images, and is more pronounced for UNAM-KIAS \citep{Nair2010,Oh2012,Masters2012,Erwin2018,Mukundan2025}. A plausible reason is the reduction in bar size \textit{relative} to the galaxy with minima at an intermediate mass, making it difficult to identify bars
as the mass approaches the turn-over value (see \citealt{Erwin2018} for a discussion). At higher masses, both galaxy and bar sizes increase with mass, and so does the likelihood of identifying the bar. Additionally, the higher bar fractions at lower masses may be driven
by the fact that fainter galaxies are located at lower redshifts and are therefore resolved better.

We find that the global bar fractions in UNAM-KIAS are consistent with recent measurements based on visual inspection of SDSS images \citep[e.g.][]{Lee2019}, but AMIGA shows lower fractions at $\mstar\gtrsim 10^{10.2}\,\msol$. One may hypothesize that this is related to the higher $\HI$ fractions in AMIGA (Fig.~\ref{hifracvsmstar}), because the dynamical support from $\HI$ helps stabilise the galaxy against instabilities and bar formation. However, we find that the offset in $\HI$ fraction vanishes when we include only galaxies
with bar classifications (not shown). Moreover, the differences in the $\HI$ fraction do not explain the inverse systematic for the strongly-barred systems, where UNAM-KIAS has a preference for slightly \textit{lower} fractions. Hence, $\HI$ content may not be the main factor in the formation of bars in isolated galaxies. UNAM-KIAS is broadly consistent with the literature on the strong bar fractions  within reasonable uncertainties \citep[e.g.][]{Masters2012,Melvin2014,Lee2019,Aquino-Ortiz2025}, but AMIGA shows slightly higher values.

It is tempting to attribute the similarity between the bar fractions in UNAM-KIAS and in the general population of disky galaxies to the weaker isolation strategy in this sample compared to AMIGA (Section~\ref{amiga}). We know, however, that UNAM-KIAS galaxies are significantly more isolated than normal spirals; this is evidenced by their high $\HI$
fractions, for example. Hence, the above results favour a scenario in which tidal interactions between galaxies have minimal contribution to the formation of bars in most cases, as claimed by recent works based on both simulations \citep{Frosst2025,Zheng2025} and observations \citep{Aquino-Ortiz2025}. In fact,
the net effect of such interactions could be the destruction of bars instead, especially when they are associated with major mergers \citep[e.g.][]{Li2026}. 

The situation is less clear for AMIGA, where the lower global bar fractions suggest a greater degree of isolation than in UNAM-KIAS, but the higher incidence of strong bars contradicts this picture. We caution against over-interpretation here, as our AMIGA sample covers only a small fraction of the parent AMIGA sample. Additional analysis with more representative samples is required to overcome this caveat and obtain further clarity, which is beyond the scope of this paper.

\begin{figure}
    \centering
    \includegraphics[width=1\linewidth]{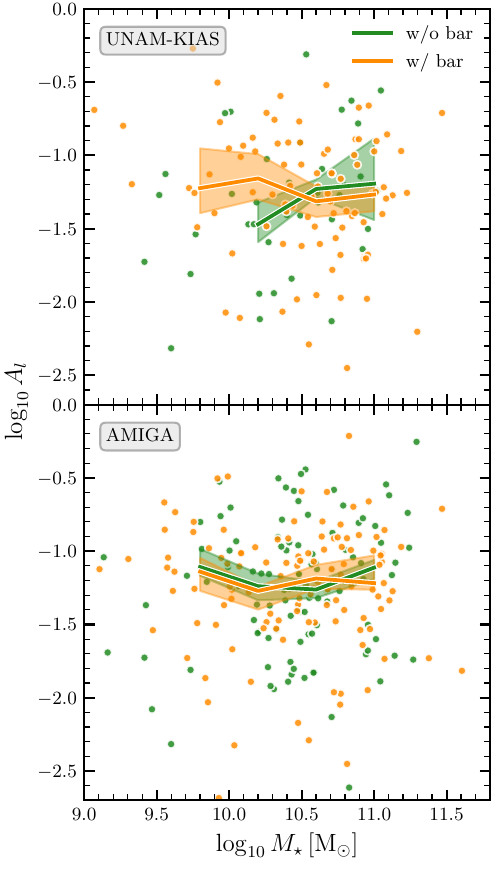}
    \caption{$\HI$ lopsidedness for galaxies with and without bars. The asymmetry is agnostic to bar presence for all masses.}
    \label{alandbar}
\end{figure}

We now proceed to explore the connection between bar presence and $\HI$ line asymmetry. Fig.~\ref{alandbar} shows the asymmetries for the barred (orange) and unbarred galaxies (green) in separate panels for UNAM-KIAS and AMIGA. Again, we plot the medians only for bins with at least 8 points. The asymmetry is clearly agnostic to bar presence for both samples, and we find similar consistency between strongly-barred and unbarred objects as well. We find these results to be robust against inclination effects. 

The lack of any link between bar and $\HI$ asymmetry could occur if the $\HI$ in our galaxies is too extended to develop \textit{global} perturbation modes. This is because the bar's impact is generally limited to the gas within its co-rotation radius that extends only to a few kpc \citep{Fanali2015,Khoperskov2018,Verwilghen2024}. This radius is significantly smaller than the $\HI$ radii of $\gtrsim 15$~kpc inferred for our galaxies from the $\HI$ size--mass relation \citep{Wang2016}. 

\subsection{Ongoing/past mergers}
UNAM-KIAS and AMIGA aim to exclude strong perturbers, but the absence of perturbers at the time of observation does not rule out past interactions \citep[see][]{Hirschmann2013}. This is important because galaxies can retain morphological and kinematical disturbances for $\approx 1$~Gyr after their last encounter \citep{Bickley2021}. It is difficult to identify such cases just from
their local environment, particularly for the so-called backsplash galaxies: the satellites that have undergone one or more pericentric passages around a massive galaxy but currently appear as isolated centrals because they are near the apocentres of their elongated orbits (see \citealt{Ruiz2023} and the references therein). Furthermore, simulations show that close interactions with minor companions can also produce notable responses in the galaxy \citep{Lipnicky2018,Ghosh2022}. For these reasons, it is crucial to investigate the extent to which such past or ongoing interactions bias the inferred $\HI$ asymmetries of our galaxies.

To address this, we perform merger classifications using the machine learning algorithm by \citet{Nevin2023} and the $r$-band images. The classification begins by flagging the galaxy as a `major merger', `minor merger', or `no merger' -- accounting for both past and ongoing mergers. Then, if the galaxy is flagged as a merger, we categorise it further as `pre-' or `post-coalescence' (for details, see Section~\ref{merger}). 

\begin{figure}
    \centering
    \includegraphics[width=1\linewidth]{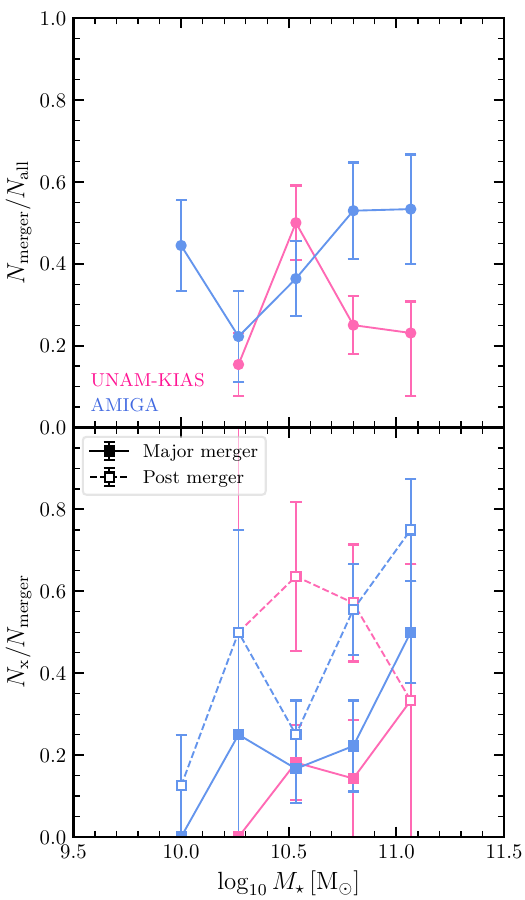}
    \caption{Merger statistics for our samples. The top panel shows the fraction of galaxies involved in past or ongoing mergers. The bottom panel shows the fraction of mergers that are major (filled squares) and those that correspond to the post-coalescence stage (open squares).
    UNAM-KIAS and AMIGA are shown as pink and blue, respectively. Among our galaxies, $\lesssim 50$ per cent have been affected by mergers, but these were usually minor mergers that occurred in the past, akin to isolated galaxies in simulations (see the text).}
    \label{fmergervsmstar}
\end{figure}

Fig.~\ref{fmergervsmstar} shows the merger statistics across stellar mass bins. These statistics are computed only for galaxies with merger information. We only plot the values for the bins with more than 10 objects. The top panel shows the fraction of galaxies affected by mergers. It seems that mergers have impacted $\sim 15-50$ per cent of our samples at any given mass. The merger fractions of $\lesssim 0.5$ broadly align with the qualitative expectation for galaxies located in low-density regions. At first glance, this also seems to be consistent with simulations based on $\Lambda$CDM, which predict 45 per cent of isolated galaxies to experience one or more mergers throughout history \citep{Hirschmann2013}. However, given that mergers taking place in early history would not show any remnant signatures and evade detection by our classification scheme, the relevant quantity for comparing against theoretical predictions is the fraction of galaxies that underwent mergers in the last Gyr or so. \citet{Hirschmann2013} predict this statistic to be 5 per cent, which is significantly below the minimum value of $\sim 15$ per cent in Fig.~\ref{fmergervsmstar}.

The bottom panel of Fig.~\ref{fmergervsmstar} provides additional insights into the nature of these mergers, as it plots the fraction of mergers characterised as major mergers and those in the post-coalescence stage. We find that most of these mergers are of minor type (filled squares) and tend to be post-coalescence (open squares). Since both UNAM-KIAS and AMIGA were specifically designed to minimise the likelihood of ongoing major mergers, these results indicate that the isolation criteria adopted for constructing these samples are effective in achieving their desired goals. In addition, this is consistent with the theoretical expectations for isolated galaxies by \citet{Hirschmann2013}.

The high merger fractions for our galaxies are rather intriguing, given that these are supposed to be isolated. We find that the fraction reported by \citet{Hirschmann2013} is based on the two-dimensional isolation criteria adopted for AMIGA by \citet{Verley2007a}, which rules out differences in isolation as the underlying cause. \citet{Hirschmann2013} also predict the merger fractions for non-isolated galaxies to be the same as those for isolated ones, meaning that our merger fractions are also higher than those expected for normal galaxies. This is a strong indication of a potential sampling bias at play, and is rather probable considering that we do not possess merger classifications for about half of our AMIGA galaxies, and our sample includes $\lesssim 10$ per cent of the parent UNAM-KIAS and AMIGA samples. Additionally, note that the training datasets used by \citet{Nevin2019,Nevin2021,Nevin2023} did not include simulations where galaxy encounters do not lead to eventual mergers (i.e. flybys), meaning that the method has not been trained to distinguish between flybys and mergers, both of which produce similar morphological features. This is important for interpreting the merger fractions in Fig.~\ref{fmergervsmstar} because many of our galaxies flagged as mergers may have actually undergone flybys. In fact, we find that our merger fractions are similar to the fraction of galaxies with tidal features reported in the literature (see \citealt{Khalid2024} and the references therein). A proper examination for consistency with theory would require further analysis that is beyond the scope of this paper.

\begin{figure}
    \centering
    \includegraphics[width=1\linewidth]{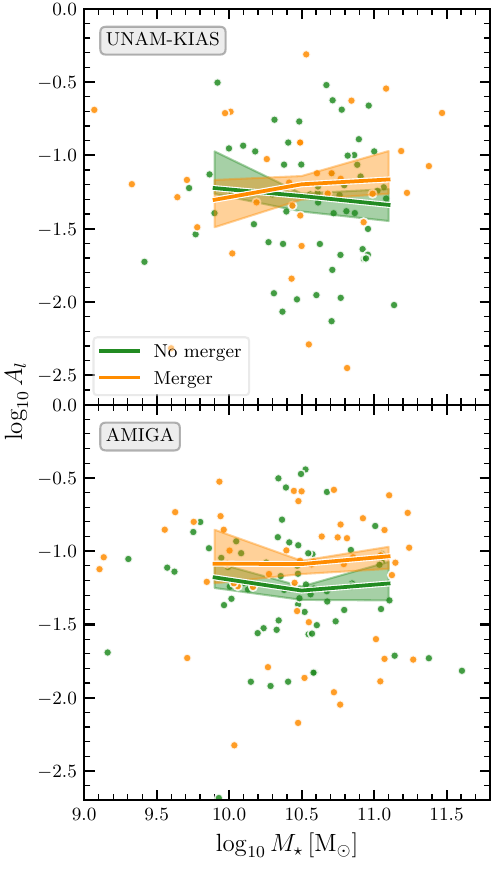}
    \caption{Dependence of $\HI$ lopsidedness on mergers. The orange points show the galaxies that are probably affected
    by a merger (minor or major), and the green points are the galaxies do not
    appear to be affected by any merger. The asymmetry is independent of mergers
    in general.}
    \label{alandmerger}
\end{figure}

Next, we address the link between mergers and $\HI$ line asymmetry in Fig.~\ref{alandmerger} that shows the asymmetries for two subsamples distinguished by colours: a) those that show no signs of mergers (green), and b) those that show signs of a merger (orange). The median values for different mass bins are shown as curves with the same colour correspondence. The results for UNAM-KIAS and AMIGA are shown in the top and bottom panels, respectively. The median is plotted only if there are $\geq 8$ points in the bin. The median asymmetries are statistically identical between the two subsamples, indicating that mergers do not significantly impact $\HI$ line asymmetry.

The broad lack of trend between mergers and asymmetry in Fig.~\ref{alandmerger} presents an interesting puzzle because a galaxy must show visible signs of interactions in its stellar disk for the algorithm to tag it as a minor merger in the first place. Simulations suggest that the event must have happened within 850 Myr or so for these disturbances to persist until the time of observation  \citep{Ghosh2022}. This is not long enough for the perturbation modes in $\HI$ to settle completely, but it is sufficient to abate them substantially \citep[e.g.][]{Chakrabarti2011}. However, we note that we do not have appreciable statistics here, mainly because many of our galaxies lack merger information. This can potentially preclude the trend from manifesting even if it is present in the full sample. In fact, this could be the reason why the sole (albeit small) offset in Fig.~\ref{alandmerger} appears for AMIGA at $\mstar\sim 10^{10.5}\,\msol$, which happens to be the only mass bin with more than 20 galaxies. We can address this better if all our galaxies have merger classifications, and defer this for a future study.  

Since major mergers cause a stronger reconfiguration of the phase space than minor ones \citep[e.g.][]{Lotz2010}, there is a greater likelihood of observing a correlation with asymmetry if we restrict our merger sample to major mergers. We could
have tested this for our galaxies, but the limited statistics again prevent
us from deriving any meaningful conclusions by binning the data as in Fig.~\ref{alandmerger}. We nevertheless conducted a KS test to examine global offsets but did not find any. The same holds for the comparison between the merger stages (post- or pre-coalescence).

\section{Comparison against previous samples}\label{previous}
Some samples of isolated $\HI$ sources have already been presented before this work. The most noteworthy of these are by \citet{Haynes1998}, \citet{Matthews1998}, and \citet{Espada2011}. \citet{Espada2011}, in particular, has been
used frequently as the standard
control sample when exploring the impact of environment on $\HI$ line asymmetry \citep[e.g.][]{Scott2018,Bok2019,Reynolds2020,Watts2020}. In this section,
we compare the asymmetries in our samples against these prior releases.

Before we perform these comparisons, it must be ascertained that the asymmetries are homogeneous across data sets. We do so by performing consistency checks for four aspects: the determination of profile edges, the asymmetry measure, the decimal precision, and the $S/N$ of the spectrum. 

We note that, unlike this study, the profile edges were derived in the literature using the `peak method' that involves identifying the velocities where the flux drops to 20 per cent of the peak flux. For the analysis in this section, we use the asymmetries computed via this method. Also, none of the other studies except \citet{Matthews1998} had employed $A_l$. \citet{Haynes1998} and \citet{Espada2011} quantified the asymmetry as the flux ratio ($A_{fr}$), which is the ratio of the integrated flux on the lower and higher velocity sides of the profile. We convert the $A_{fr}$ in \citet{Haynes1998} and \citet{Espada2011} to $A_l$ as
\begin{equation}
A_l = \frac{|A_{fr}-1|}{A_{fr}+1}.    
\end{equation} 
Likewise, we round off the asymmetries in our samples to match the precision of the sample they are compared against. Specifically, this involves rounding off to two decimal places when comparing with \citet{Haynes1998} and \citet{Matthews1998}, and to three decimal places in the case of \citet{Espada2011}. This avoids spurious deviations due to truncation. In addition, we examine the $S/N$ within these samples and find that \citet{Haynes1998} and \citet{Espada2011} show $S/N\gtrsim 10$, but $\approx 47$ per cent of the \citet{Matthews1998} sample lies below this threshold. For this reason, we only compare with the \citet{Matthews1998} subsample above $S/N=10$.

\begin{figure}
    \centering
    \includegraphics[width=0.8\linewidth]{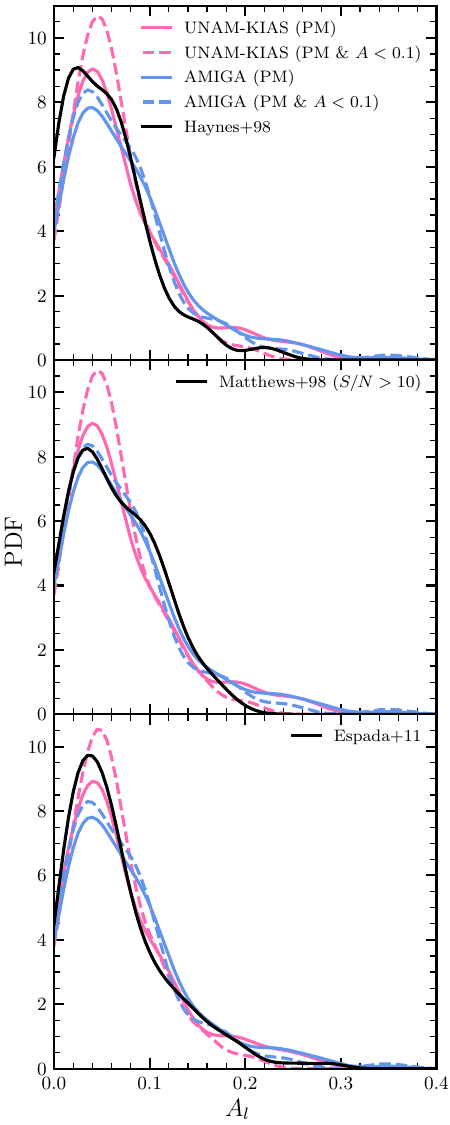}
    \caption{$\HI$ line asymmetries in UNAM-KIAS and AMIGA compared against those in previously published samples of isolated galaxies in the literature: \citet{Haynes1998} (top),  \citet{Matthews1998} (middle), and \citet{Espada2011} (bottom). These samples are shown as the black curves. The solid coloured curves show the asymmetries for our samples based on the peak method, which is the method adopted in the literature (see the text). The dashed coloured curves are the asymmetries for the galaxies with low levels of stellar perturbation ($A<0.1$).
    }
    \label{comparewprev}
\end{figure}

The results are shown in Fig.~\ref{comparewprev}, where each panel is dedicated to comparison with one of the three samples from the literature, corresponding to \citet{Haynes1998}, \citet{Matthews1998}, and \citet{Espada2011} from top to bottom. The top panel shows that UNAM-KIAS is closer to \citet{Haynes1998} than AMIGA. This is supported by our KS tests that show UNAM-KIAS to be statistically equivalent, but show a discrepancy with AMIGA ($p$-value $\approx 0.006$). Apart from this, the only discrepancy we find at $>2\sigma$ level ($p$-value~$<0.05$) is between AMIGA and \citet{Espada2011} (bottom panel). Our Mood's tests show that the median asymmetries do not exhibit any notable difference, which is understandable given that the peaks occur at similar locations. This means that the above discrepancies with AMIGA primarily arise due to the tail extending towards high asymmetries above $A_l\approx 0.08$.

At this point, it is important to consider whether isolation was the sole criterion for building the aforementioned samples. \citet{Matthews1998} only included galaxies that appear regular in their optical images \citep{Matthews1997}, which is expected to introduce some bias towards lower asymmetries for all the baryonic components within the galaxies -- even though the global correlation between stellar and $\HI$ asymmetry seems rather modest (Fig.~\ref{alvsas}). Our samples show lower asymmetries when we restrict them to systems with low degrees of stellar perturbation by imposing $A<0.1$ (dashed curves). In fact, the discrepancies seen for AMIGA with \citet{Haynes1998} and \citet{Espada2011} alleviate to $\lesssim 2\sigma$ significance with the new subsample. We therefore speculate that the samples used by these authors have a preference for
isolated galaxies that appear visually unperturbed, although this is not explicitly stated in the papers. In this context, it appears that our samples contain higher fractions of asymmetric galaxies than prior samples because they span a wider diversity of isolated galaxies in the local Universe. Moreover, recall that our samples preferentially include galaxies that have near-complete $\HI$ coverage by the telescope beam (Sections~\ref{cross} \&~\ref{beam}). This is important because perturbed structures in extended $\HI$ disks are often located at the outskirts \citep[e.g.][]{Sengupta2012,Scott2014,Zheng2022} and can evade the beam for galaxies at small distances. Nevertheless, readers can select `purer' subsamples using stellar asymmetries and merger information, if needed.

\section{Summary and conclusions}\label{summary}
In this study, we presented the isolated galaxy samples from UNAM-KIAS \citep{Hernandez-Toledo2010} and AMIGA \citep{Argudo-Fernandez2015} with robust $\HI$ line asymmetry measurements [$A_l$; equation~(\ref{lop})] based on the global (spatially-unresolved) 21-cm spectra from
$\HI$-MaNGA \citep{Masters2019,Stark2021}, NIBLES \citep{vanDriel2016}, KLUN \citep{Theureau2005,Theureau2007,Theureau2017}, ALFALFA \citep{Haynes2018}, or xGASS \citep{Catinella2018}. These samples benefit from wide sky coverage, rigorous isolation criteria, near-complete beam coverage, and minimal contamination from source confusion, instrumental noise, and poor velocity resolution (Section~\ref{method}). To guide future analyses, we discussed the biases associated with overlap, inclination, stellar mass, morphology, stellar asymmetry, star formation rate, $\HI$ mass, halo mass, bar presence, and mergers (Section~\ref{results}). We additionally compared our samples with prior isolated $\HI$ samples in the literature (Section~\ref{previous}).

Whilst there are some differences between UNAM-KIAS and AMIGA, they also present consistencies that underscore the aspects that hold true for isolated galaxies. To summarise, these galaxies tend to be star-forming (Fig.~\ref{ssfrvsmstar}) and LTC (Fig.~\ref{morphvsmstar}), and exhibit lower star formation efficiencies (Fig.~\ref{sfevsmstar}) and higher $\HI$ fractions (Fig.~\ref{hifracvsmstar}) than is typical for such objects -- where the latter is likely caused by weaker outflows and stronger gas accretion (at $\mstar\lesssim 10^{10.3}\,\msol$) in addition to lower star formation efficiencies (see Section~\ref{hifractions}). These findings align with theoretical works that predict isolated galaxies to exist in the quasi-equilibrium regime where star formation is self-regulated by the turbulence from feedback, maintaining the Toomre parameter $Q\sim 1-2$ \citep[e.g.][]{Firmani-Avila2000,Hopkins2011,Hopkins2012,Becerra2014}. Likewise, the bar fractions tend to be similar to the general population of disky galaxies (Fig.~\ref{fbarvsmstar}), providing support for the idea that bar formation is primarily a secular process, as suggested by recent theoretical and observational studies \citep{Aquino-Ortiz2025,Frosst2025,Zheng2025}. 

The recent merger histories of our isolated galaxies seem rather peculiar because $\sim 15-50$ per cent of them have been affected by mergers at any given mass. These fractions are significantly higher than the theoretical prediction of 5 per cent for mergers in the last Gyr, for both isolated and non-isolated populations \citep{Hirschmann2013}. The dynamic range is, however, similar to that of the fraction of galaxies with tidal features \citep[see][]{Khalid2024}, and these features can be produced during interactions that do not lead to eventual mergers (flybys). The unusually high merger fractions of our galaxies could be due to sampling bias and/or the inability of the classification method to distinguish between flybys and encounters that lead to coalescence (Section~\ref{merger}). The precise reason remains uncertain.

With regard to the $\HI$ line asymmetry, the general consensus between the samples is that it correlates positively with sSFR at $\mstar\lesssim 10^{10.3}\,\msol$, which is in line with the predictions for central galaxies from the \textsc{\large eagle} simulation \citep{Manuwal2022} and suggests
stellar feedback as a possible modulator of its scatter in this mass range (see Section~\ref{ssfr}). The two also show agreement in the independence of asymmetry on stellar mass (Fig.~\ref{alvsmstar}), shape asymmetry ($A_s$; Fig.~\ref{alvsas}), morphology (Fig~\ref{alvsmorph}), bar presence (Fig.~\ref{alandbar}), $\HI$ fraction (Fig.~\ref{hifracandal}), and mergers (Fig.~\ref{alandmerger}).

A few additional, noteworthy results are stated below:
\begin{enumerate}
    \item Each sample exhibits a trend with $\HI$ line asymmetry that is specific to the sample: UNAM-KIAS shows a positive correlation with inclination (Fig.~\ref{alvsinc}), whereas AMIGA shows it with rotational stellar asymmetry ($A$; Fig.~\ref{alvsas}). 
    
    \item UNAM-KIAS shows systematically lower $\HI$ masses relative to AMIGA (Fig.~\ref{hifracvsmstar}). Given that the samples present consistent star formation efficiencies, sSFRs, and halo masses (Fig.~\ref{shmr}) at fixed $\mstar$, and AMIGA employs a stricter isolation strategy (Section~\ref{amiga}), we conclude that the offset in $\HI$ mass reflects the higher local densities probed by UNAM-KIAS, which is expected to reduce the accretion rate of gas onto the galaxy (Section~\ref{hifractions}). 

    \item Our UNAM-KIAS sample is statistically indistinguishable from prior samples of isolated galaxies in the literature (Fig.~\ref{comparewprev}), but AMIGA differs from \citet{Haynes1998} and \citet{Espada2011} samples due to a larger proportion
    of asymmetric galaxies (Fig.~\ref{comparewprev}). These discrepancies vanish if AMIGA is restricted to the galaxies with low stellar perturbations, suggesting that our samples span a wider diversity of isolated galaxies.
\end{enumerate}   

The samples presented in this work can be improved and extended further by tapping into the full potential of UNAM-KIAS and AMIGA. Radio facilities like the SKA can provide 21-cm spectra at greater $S/N$ and spatial resolution. This could enable the inclusion of galaxies that were excluded due to our quality cuts (Sections~\ref{beam},~\ref{sbyn} \&~\ref{chanandmag}), and allow reliable measurements for other asymmetry measures that are more sensitive to noise than $A_l$; these can be combined with $A_l$ to yield a measure with more information than $A_l$ alone \citep[e.g.][]{Manuwal2022}. Additionally, resolved $\HI$ observations would help in gauging the contributions from kinematic and spatial asymmetry, and aid in better elucidation of the physics underlying the global asymmetries of isolated galaxies.

We hope that our isolated $\HI$ samples and the aforementioned efforts would aid in extracting novel insights into the origin of $\HI$ line asymmetry, and improve our understanding of its role within the broader context of $\HI$ and galaxy evolution.
 
\section*{Acknowledgements}
We would like to thank the anonymous referee for a constructive review that was conducive to improvements in key aspects of this paper. AM acknowledges support through a postdoctoral fellowship from The Secretariat of Science, Humanities, Technology and Innovation (SECIHTI). The authors acknowledge financial support from SECIHTI (former CONAHCyT) project CF-G-543 and DGAPA-PAPIIT IN106823. HMHT also acknowledges support from the SECIHTI project CF-2023-G-1052.

The following \textsc{\large python} packages were employed in this work: \textsc{\large matplotlib} \citep{Matplotlib}, \textsc{\large numpy} \citep{Numpy}, \textsc{\large scipy} \citep{Scipy}, and \textsc{\large astropy} \citep{Astropy}. The paper has been typeset using Overleaf\footnote{\url{https://www.overleaf.com/}}.

%%%%%%%%%%%%%%%%%%%%%%%%%%%%%%%%%%%%%%%%%%%%%%%%%%
\section*{Data Availability}
The $\HI$ spectra are available through the respective links. $\HI$ line asymmetries and other ancillary
properties are provided online as supplementary material.


\bibliographystyle{mnras}
\bibliography{references} % if your bibtex file is called example.bib

%%%%%%%%%%%%%%%%%%%%%%%%%%%%%%%%%%%%%%%%%%%%%%%%%%

%%%%%%%%%%%%%%%%% APPENDICES %%%%%%%%%%%%%%%%%%%%%

\appendix
\section{Asymmetry Convergence}\label{conv}
Since the asymmetries of $\HI$ spectra with low $S/N$ are artificially inflated, a high incidence of such emission lines in the sample would pose a problem for proper interpretation of trends. One could, of course, enforce an extremely large $S/N$ threshold (e.g. $S/N>100$), but this is impractical for 
currently available $\HI$ data, as there would not be enough statistics to derive any meaningful conclusions. It is more reasonable to use a threshold that is sufficiently high to enable robust asymmetry measurements for a large proportion of the sample, and also provides appreciable statistics for carrying out the science.

\begin{figure}
    \centering
    \includegraphics[width=1\linewidth]{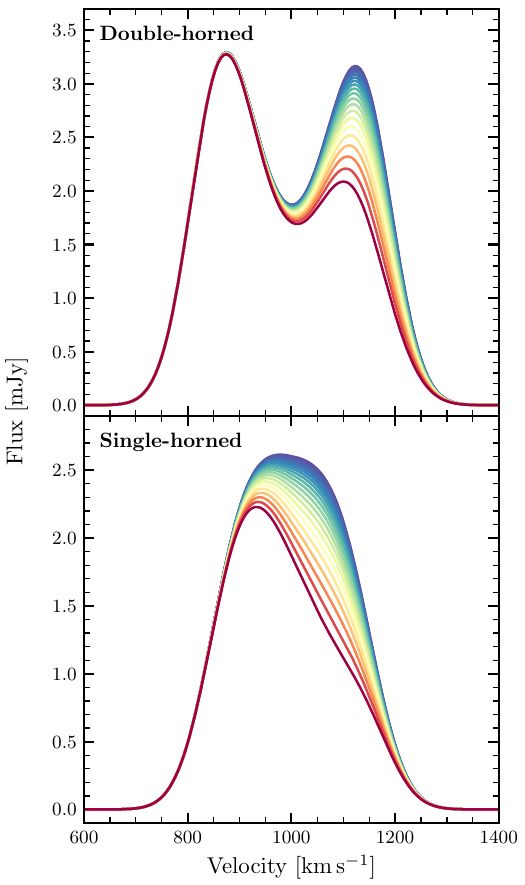}
    \caption{The model $\HI$ lines used for convergence tests for the cases when the profile is double-horned (top) and single-horned (bottom). The colour of the profile indicates its asymmetry, increasing from blue to red.}
    \label{models}
\end{figure}

We conduct extensive tests using model $\HI$ profiles to explore the relationship between $S/N$ and line asymmetry. This is carried out for the two types of profiles broadly presented by our spectra: double-horned and single-horned. The first represents highly inclined disks, and the second one indicates face-on disks or dispersion-supported systems. For each type, we generate 20 noiseless variations with approximately the same profile width but different degrees of asymmetry. In detail, these asymmetries are implemented by damping the right half of the perfectly symmetric spectrum. We take the velocity resolution $v_{\rm res} = 5.5\,\kms$, as a vast majority of our matches are from ALFALFA. The profiles are displayed in Fig.~\ref{models}, where redder profiles exhibit greater levels of asymmetry. Once we have generated the noiseless profiles, we add varying levels of Gaussian noise to each, generate 1000 realisations per noise level, and compute the corresponding $S/N$ values and the three asymmetry measures employed by \citet{Manuwal2022}: i.e. lopsidedness ($A_l$), velocity offset ($A_{vo}$), and normalised residual ($A_{nr}$). This provides good statistics at different $S/N$ values for each measure and allows us to capture its convergence with $S/N$ accurately. All the asymmetries are computed based on the profile edges derived from the noiseless version, similar to the methodology for real spectra where the underlying noiseless profile is approximated using the best-fit Busy function (Section~\ref{sbyn}).

\begin{figure*}
    \centering
    \includegraphics[width=1\linewidth]{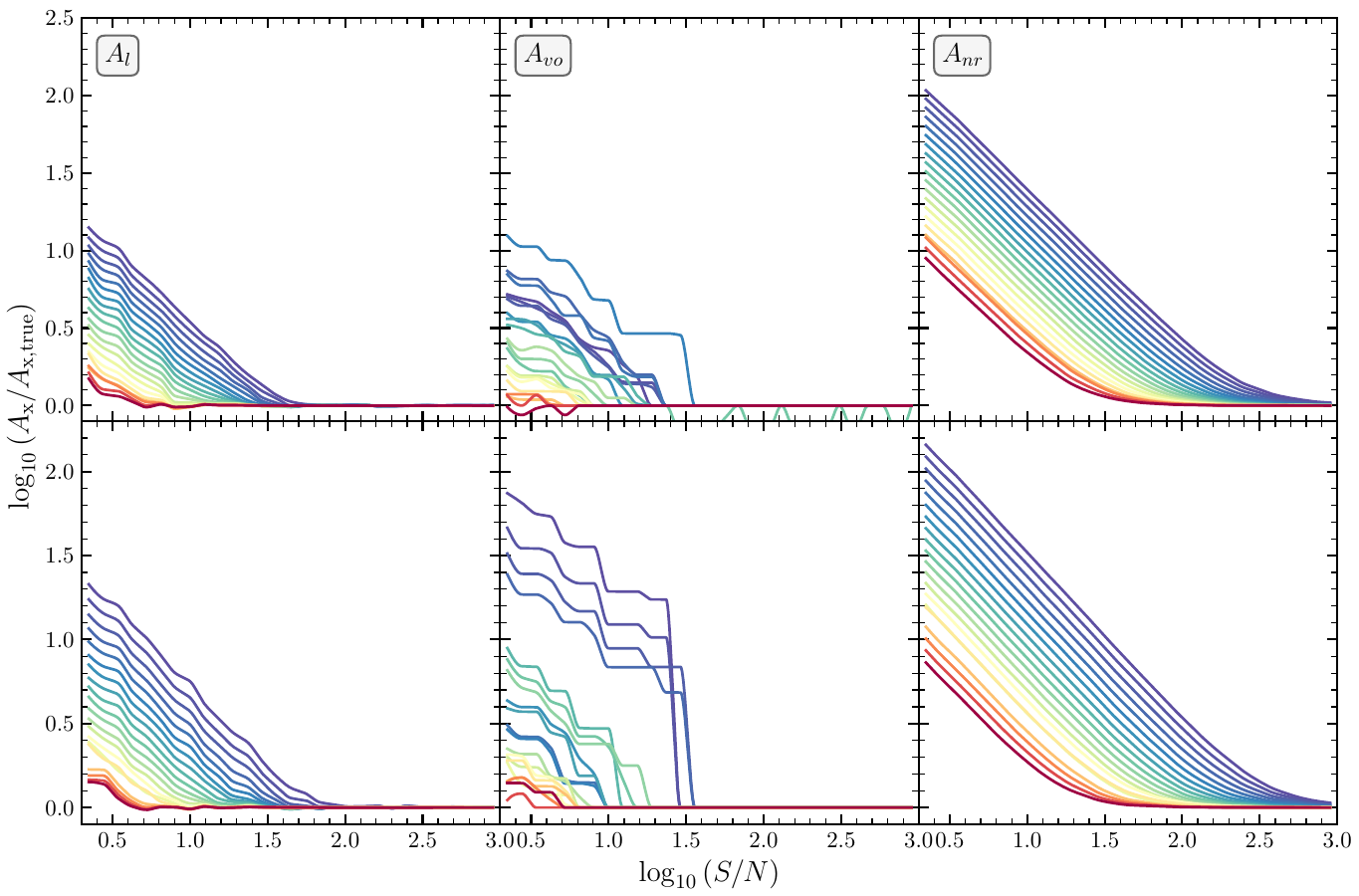}
    \caption{The dependence of line asymmetry on $S/N$ for the model $\HI$ profiles in Fig.~\ref{models}. {\rm The top and bottom rows show the results for double-horned and single horned profiles, respectively. In each row, the} panels from left to right show the results for $A_l$, $A_{vo}$, and $A_{nr}$, where the vertical axis plots the ratio of the measured asymmetry to the true (noiseless) asymmetry for the measure under consideration (mentioned in the top-left corner of the panels in the top row).
    The colours correspond to those in Fig.~\ref{models}. For both profile types, profiles with higher degrees of intrinsic asymmetry tend to exhibit convergence at lower $S/N$, although the trend is not always monotonic for $A_{vo}$. Also, $A_{nr}$ requires higher $S/N$ to converge than the other two measures.}
    \label{asymvssbyn}
\end{figure*}

The results from this exercise are shown in Fig.~\ref{asymvssbyn}, where the top and bottom rows correspond to double-horned and single-horned profiles, respectively, and the panels in a given column plot the median ratio of the measured asymmetry ($A_{\rm x}$) to the true value (for the noiseless profile; $A_{\rm x,\,true}$) against $S/N$ for the measure stated in the top row. The colours correspond to the models in Fig.~\ref{models}. The curves have been smoothed at 1000 evenly-spaced $\log_{10} S/N$ values using a shape-preserving interpolator to improve the sampling in regions with rapid changes in $A_{\rm x}/A_{\rm x,\,true}$, $A_{vo}$ in particular.

There are three main inferences that we derive from these curves. First, $A_{nr}$ always converges at significantly higher $S/N$ values than the other two measures. This occurs due to the channel-by-channel differences incorporated in this measure that organically render it more sensitive to instrumental noise. More importantly, this means that even the combined measure ($A_{\rm comb}$) in \citet{Manuwal2022}, which is the arithmetic mean of the three measures, would converge at a $S/N$ higher than that required for $A_l$'s or $A_{vo}$'s convergence. Therefore, although $A_{\rm comb}$ encapsulates more information about the profile's asymmetry than any individual measure, it should only be used if the spectral sample at hand exhibits remarkably high $S/N$ values. The second important inference is that the profiles with greater $A_{\rm x,\,true}$ generally converge at lower $S/N$ values. We attribute this to the fact that the relative deviation from true asymmetry due to the added noise is greater for symmetric lines. In other words, adding noise to a highly asymmetric line changes its asymmetry by a smaller fraction because it is already asymmetric to begin with. Lastly, we find that this behaviour is not followed strictly by $A_{vo}$, which sometimes shows convergence at higher $S/N$ for profiles that are visually {\it more} asymmetric.

\begin{figure}
    \centering
    \includegraphics[width=1\linewidth]{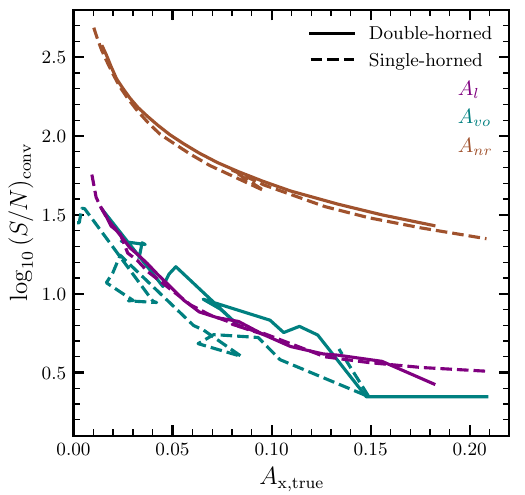}
    \caption{The $S/N$ threshold required for convergence within 20 per cent for profiles with varying levels of intrinsic (noiseless) asymmetry. The colour depicts a specific asymmetry measure as per the legend in the top-right. The solid and dashed curves correspond to double-horned and single-horned profiles, respectively. The threshold shows a monotonic decrease with asymmetry for $A_l$ and $A_{nr}$, whereas $A_{vo}$ shows an erratic behaviour. $A_{vo}$ also requires higher thresholds for double-horned profiles, whereas $A_l$ and $A_{nr}$ show the same convergence for both profile types. $A_{nr}$ always requires higher convergence thresholds than those for the other measures by $\gtrsim 1$~dex.}
    \label{sbynconv}
\end{figure}

For better quantitative clarity, we determine the $S/N$ thresholds required for convergence for different values of $A_{\rm x,\,true}$. To do so, we identify the $S/N$ at which the measurements agree within 20 per cent of $A_{\rm x,\,true}$, $(S/N)_{\rm conv}$, taken as the minimum $S/N$ where $|A_{\rm x}/A_{\rm x,\,true}-1|<0.2$. Fig.~\ref{sbynconv} shows $(S/N)_{\rm conv}$ for different values of $A_{\rm x,\,true}$ for the three measures that are depicted using different colours. For each measure, we show a solid curve for the double-horned profiles and a dashed curve for the single-horned profiles. The figure shows that $(S/N)_{\rm conv}$ broadly increases with $A_{x,\,{\rm true}}$ and is particularly high for exquisitely symmetric lines, reaching $\approx 500$ for $A_{nr}$. The curve for $A_{vo}$ stands out and is rather erratic. To understand this, it is important to note that the curves are plotted following the same rank order of the profiles, that is, the values are plotted from least to most asymmetric in Fig.~\ref{models}. Hence, the loops in the curve for $A_{vo}$ show that this measure does not always increase as the profiles become qualitatively more asymmetric. We have verified that this non-monotonicity occurs even when the profile edges have not changed, meaning that this is an inherent drawback of this measure. An additional way in which $A_{vo}$ differs from the other measures is that the double-horned profiles with intermediate asymmetries require larger $S/N$ thresholds for convergence than the single-horned profiles.

Our tests clearly indicate that $A_{nr}$ requires about 10 times higher $S/N$ thresholds than other measures, $A_{vo}$ does not always capture the asymmetry as expected, and the threshold $S/N$ for its convergence depends on the profile type (double-horned vs single-horned). For these reasons, we prefer to avoid these measures in this study and only use $A_l$. However, Fig.~\ref{sbynconv} suggests that robust $A_l$ measurements over the full range of profiles explored here demand imposing a threshold of $S/N\approx 60$. This cut is too high for constructing a sample with appreciable statistics, implying that a lower $S/N$ cut must be adopted to enable useful analyses. We adopt a fiducial threshold of $S/N=10$ to select our galaxies because it seems to provide an appropriate balance between measurement accuracy and statistics. This allows decent convergence for $A_l\gtrsim 0.05$ and is a reasonable choice because galaxies with lower asymmetries are not predicted to dominate the sample \citep[e.g.][]{Watts2020b,Manuwal2022}.

Note that the above tests are based on profiles with a specific $v_{\rm res}$ and without any smoothing, and the real spectra demonstrate inhomogeneity in both of these properties (see Section~\ref{hisamples}). Whether this is an important caveat of our conclusions above depends on the degree to which the $(S/N)_{\rm conv}$ vs $A_{l,\,\rm true}$ curve (like those in Fig.~\ref{sbynconv}) is agnostic to $v_{\rm res}$ and $v_{\rm eff}$. To examine this, we do additional tests with different values of these two properties, chosen to be representative of the parent $\HI$ samples considered in this work. We use $v_{\rm res}=1.2,1.4,2.6,5.5\,\kms$ and the smoothing involves convolutions with the boxcar kernel for $v_{\rm eff}=v_{\rm res}$ (equivalent to no smoothing) in addition to one or more $v_{\rm eff}>v_{\rm res}$. In detail, this includes $v_{\rm eff}=10\,\kms$ for $v_{\rm res}=1.2$ ($\HI$-MaNGA) and $5.5\,\kms$ (ALFALFA), $v_{\rm eff}=5,10,15\,\kms$ for $v_{\rm res}=1.4\,\kms$ (xGASS), and $v_{\rm eff}=10$ (KLUN) and $18\,\kms$ (NIBLES) for $v_{\rm res}=2.6\,\kms$. We quantify the impact of velocity resolution and smoothing by obtaining the $(S/N)_{\rm conv}$ vs $A_{l,\,\rm true}$ curves for different $v_{\rm res}$--$v_{\rm eff}$ pairs and noting the $A_{l,\,\rm true}$ where $(S/N)_{\rm conv}=10$. These asymmetries represent the minimum $A_l$ values that can be measured reasonably accurately with this $S/N$ cut for various $v_{\rm res}$--$v_{\rm eff}$ combinations.

\begin{figure}
    \centering
    \includegraphics[width=1\linewidth]{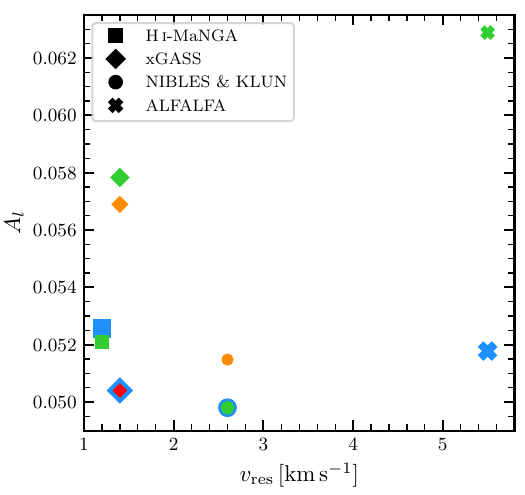}
    \caption{The impact of profile smoothing and raw velocity resolution on 
    $A_l$'s convergence. Each point shows the asymmetry corresponding to $S/N=10$
    in the $(S/N)_{\rm conv}$ vs $A_{l,\,\rm true}$ curve for a given combination of $v_{\rm res}$ and $v_{\rm eff}$. The marker style indicates a specific $v_{\rm res}$ and the colour indicates $v_{\rm eff}$ with values increasing from blue (no smoothing) to red.}
    \label{smoothtest}
\end{figure}

Fig.~\ref{smoothtest} shows the $A_l$ values derived this way, where marker styles represent different values of $v_{\rm res}$ (i.e. raw resolution), and the colours indicate different levels of smoothing, with blue corresponding to no smoothing ($v_{\rm res}=v_{\rm eff}$) and successively redder colours being higher $v_{\rm eff}$ values (more smoothing). The parent $\HI$ sample represented by the $v_{\rm res}$ is mentioned in the top-left corner. The figure suggests that asymmetries of unsmoothed profiles can be accurately measured for $A_l\gtrsim 0.05$ at each $v_{\rm res}$, but this can change depending on the smoothing level. Interestingly, the qualitative variation of $A_l$ with smoothing is not universal and instead depends on the $v_{\rm res}$. For example, consider the points corresponding to $v_{\rm eff}=10\,\kms$, shown in green for $v_{\rm res}=1.2,2.6,5.5\,\kms$ and in orange for $v_{\rm res}=1.4\,\kms$. There is virtually no change in $A_l$ at $v_{\rm res}=1.2$ and $2.6\,\kms$ compared to the blue points (unsmoothed), but there is a visible increase at the other resolutions. Furthermore, the trend between $A_l$ and $v_{\rm res}$ depends on the smoothing. This can be seen using the green and blue points that show a zig-zag and a U-shaped pattern, respectively. Notwithstanding,
the main conclusion from these results is that, for the parameters explored here, smoothing can push the convergence limit of $A_l$ to $\approx 0.06$, which is higher than the $A_l\approx 0.05$ limit for unsmoothed cases. Note, however, that these deviations are within $\approx 20$ per cent, which is also the uncertainty on the $A_l$ values plotted in Fig.~\ref{smoothtest} (by definition). Hence, variations in raw velocity resolution or smoothing among our spectra are not expected to affect the robustness of the measured asymmetries or cause any systematic. Although Fig.~\ref{smoothtest} only presents the results for double-horned profiles, we have confirmed that the results are identical for the single-horned type.

% Don't change these lines
\bsp	% typesetting comment
\label{lastpage}
\end{document}